\documentclass[sigconf]{acmart}
\usepackage{graphicx}
\usepackage{subcaption}
\usepackage{bm}

\AtBeginDocument{%
  \providecommand\BibTeX{{%
    \normalfont B\kern-0.5em{\scshape i\kern-0.25em b}\kern-0.8em\TeX}}}

\setcopyright{acmcopyright}
\copyrightyear{2022} 
\acmYear{2022} 
\setcopyright{acmlicensed}\acmConference[IUI '22]{27th International Conference on Intelligent User Interfaces}{March 22--25, 2022}{Helsinki, Finland}
\acmBooktitle{27th International Conference on Intelligent User Interfaces (IUI '22), March 22--25, 2022, Helsinki, Finland}
\acmPrice{15.00}
\acmDOI{10.1145/3490099.3511142}
\acmISBN{978-1-4503-9144-3/22/03}

\acmJournal{TOG}
\acmVolume{37}
\acmNumber{4}
\acmArticle{111}
\acmMonth{8}

\begin{document}
%\pagenumbering{gobble}

\title{Multimodal Driver Referencing: A Comparison of Pointing to Objects Inside and Outside the Vehicle}

\author{Abdul Rafey Aftab}
\email{abdul-rafey.aftab@bmw.de}
\affiliation{
  \institution{BMW Group}
  \city{Munich}
  \country{Germany}
}
\affiliation{%
  \institution{Saarland University}
  \city{Saarbrücken}
  \country{Germany}
}

\author{Michael von der Beeck}
  \email{michael.beeck@bmw.de}
\affiliation{%
  \institution{BMW Group}
  \city{Munich}
  \country{Germany}
}

\makeatletter
\let\@authorsaddresses\@empty
\makeatother
\renewcommand{\shortauthors}{Aftab and von der Beeck}

\begin{abstract}
Advanced in-cabin sensing technologies, especially vision based approaches, have tremendously progressed user interaction inside the vehicle, paving the way for new applications of natural user interaction. 
Just as humans use multiple modes to communicate with each other, we follow an approach which is characterized by simultaneously using multiple modalities to achieve natural human-machine interaction for a specific task: pointing to or glancing towards objects inside as well as outside the vehicle for deictic references. By tracking the movements of eye-gaze, head and finger, we design a multimodal fusion architecture using a deep neural network to precisely identify the driver's referencing intent. Additionally, we use a speech command as a trigger to separate each referencing event. We observe differences in driver behavior in the two pointing use cases (i.e. for inside and outside objects), especially when analyzing the preciseness of the three modalities eye, head, and finger. We conclude that there is no single modality that is solely optimal for all cases as each modality reveals certain limitations. Fusion of multiple modalities exploits the relevant characteristics of each modality, hence overcoming the case dependent limitations of each individual modality. Ultimately, we propose a method to identity whether the driver's referenced object lies inside or outside the vehicle, based on the predicted pointing direction.
\end{abstract}




\begin{teaserfigure}
    \centering
    \includegraphics[trim=0 0 0 0, clip, width=\textwidth]{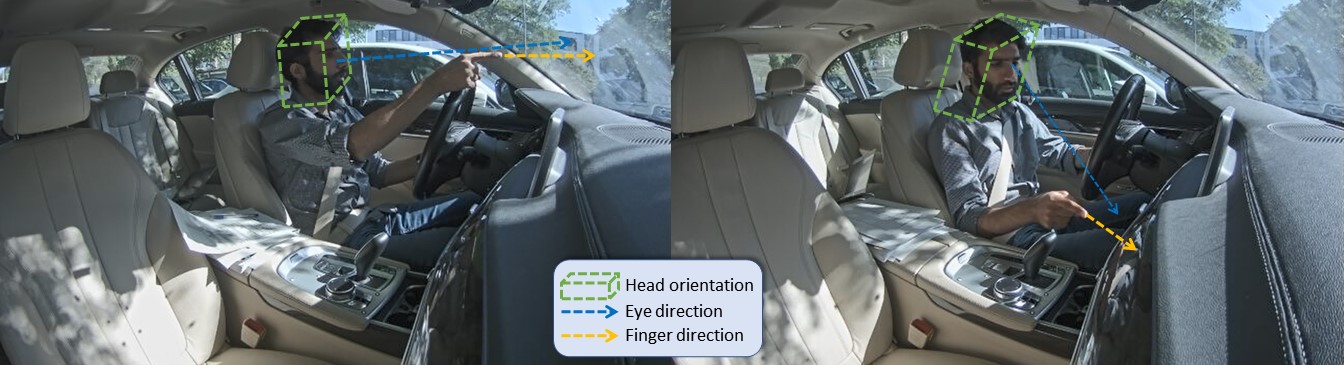}
    \caption{Left: Driver points to an outside-vehicle object. Right: Driver points to an inside-vehicle object.}
    \label{fig:teaser}
\end{teaserfigure}

\maketitle

\section{Introduction}
Interaction between the driver and the car is becoming an increasingly popular topic. Speech modality, which is commonly used in personal assistant systems in production vehicles, assists the driver by reducing touch-based interaction which can be a source of distraction \cite{tscharn2017stop, rumelin2013free}. However, using only speech as interacting modality can be cumbersome, especially in situations, where the users want to reference objects which are unknown to them. Therefore, for natural interaction involving deictic references, one needs other modalities as well, such as head pose, eye-gaze, or finger-pointing gestures. Integrating another modality for deictic referencing with speech, thereby increases usability as well as naturalness. Bolt's pioneering work, "Put that there" \cite{bolt1980put}, combined speech and gesture demonstrating the practicality of using multimodal input for natural user interaction. Multimodal interaction has since  been rigorously studied and incorporated into the car \cite{ohn2014head, nesselrath2016combining, gomaa2020studying, pfleging2012multimodal, muller2011multimodal}. 

In the context of driving experience, gesture recognition enables direct interaction with the vehicle surroundings while also allowing interaction with a wide range of in-vehicle functions. Deictic references, such as "what is \textit{that} landmark?" or "what does \textit{this} button do?," provide drivers with feedback for inquisitive commands if the referenced object can be identified. On the other hand, control commands, such as "stop over \textit{there}" or "close \textit{that} window," can assist the driver in ease of control. Though accurate detection of the driver's pointing direction is a technical challenge in itself, a more crucial problem in this task is the lack of sufficient precision in the user's pointing direction \cite{roider2018implementation, brand2016pointing}. Ray casting based finger pointing techniques  are limited by the user's pointing accuracy \cite{mayer2018effect}. Therefore, gaze input has previously been used to obtain information about the driver's focus of attention. 

In this paper, we use features from the three modalities, head pose, eye-gaze and finger pointing and fuse them to identify driver's referencing, where the referenced object may be situated inside or outside the vehicle. The drivers' behaviour while pointing to objects is also extensively studied. The contributions of this paper include: 1) a comparison between driver's pointing behavior to objects inside the car and to objects outside the car, 2) a study of precision of the three modalities, eye-gaze, head pose and finger pointing, and their fusion to identify the referenced object and 3) an effective approach to differentiate the referencing of inside-vehicle control elements and outside-vehicle landmarks. Furthermore, we discuss the importance of multimodal fusion and its effectiveness as compared to single modalities in different cases.

\section{Related Work}
Pointing gestures and eye gaze tracking as input modalities for user interaction have been studied extensively. Selection made by using gaze and head pose has been done such as by Kang et al. \cite{kang2015you}. Using gaze, even if accompanied by head pose, presents challenges particularly in cases where objects are located in close proximity to each other \cite{hild2019suggesting}. This is because gaze lacks a natural trigger due to its always-on characteristic, and has the potential to be quite volatile \cite{ahmad2017intelligent}. Use of an additional modality such as speech in addition to gaze (in order to trigger the selection of objects) has often been made. \cite{maglio2000gaze}. EyePointing \cite{schweigert2019eyepointing} makes use of finger pointing to trigger the selection of objects on a screen using gaze direction. Misu et al.\cite{misu2014situated} and Kim et al. \cite{kim2014identification} make use of head pose using speech as a trigger for driver queries of outside-vehicle objects. 

Gestures and free hand pointing add to the naturalness of the driver as they lessen the driver's cognitive demand as compared to touch based inputs, making finger pointing a useful input as well \cite{rumelin2013free, tscharn2017stop}. This was also demonstrated by Pfleging et al. in their work combining speech and gestures with minimal touch inputs on the steering wheel \cite{pfleging2012multimodal}. For selection of Points-of-Interest (POI) while driving a vehicle, Suras-Perez et al. \cite{sauras2017voge} and Fujimura et al.\cite{fujimura2013driver} used finger pointing with speech and hand-constraint finger pointing, respectively. However, the use of finger pointing can be a difficult task especially when trying to identify objects that do not lie straight ahead \cite{akkil2016accuracy}. In fact, while studying driver behaviors in a driving simulator, Gomaa et al. found gaze accuracy to be higher than pointing accuracy \cite{gomaa2020studying}.  To improve the accuracy of pointing for object selection inside the car, Roider et al. \cite{roider2018see} utilized a simple rule based approach that involves gaze tracking, and demonstrated that it improved finger pointing accuracy. However, the experiment was limited to four objects on the screen. Chaterjee et al. also combined gaze and gesture as inputs and showed better results with integration of the two than if either input is used separately \cite{chatterjee2015gaze+}. 

As research has shown, the use of multiple input modalities can surpass a single input modality in terms of performance \cite{esteban2005review, liu2018efficient, turk2014multimodal}, multimodal user interaction offers a significant utility for in-vehicle application. Mitrevska et al. demonstrate an adaptive control of in-vehicle functions using an individual modality (speech, gaze or gesture) or a combination of two or more \cite{mitrevska2015siam}. Mu\"ller and Weinberg discuss methods for a multimodal interaction using gaze, touch and speech for in-vehicle tasks presenting a few advantages and disadvantages of individual modalities \cite{muller2011multimodal}. Moniri et al. \cite{moniri2012multimodal} combined eye gaze, head pose, and pointing gestures for multimodal interaction for outside-vehicle referencing for object selection. Nesselrath et al. adopted a multimodal approach, combining three input modalities: gaze, gestures, and speech in a way that objects were first selected by gaze, e.g., windows or side mirrors, and then controlled using speech or gesture \cite{nesselrath2016combining}.

These approaches mostly use gaze information and increase the naturalness of the user interaction by including a secondary input such as speech or gesture. However, these multimodal approaches do not use the opportunity to enhance the preciseness of gaze tracking, although some use semantics from speech to narrow down the target. Our work achieves this enhancement in preciseness with the use of multimodal fusion of relevant deictic information from gaze, finger pointing, and head pose as input modalities \cite{akkil2016accuracy}. Instead of using finger pointing as a trigger for selection, we use it as an equal input modality, while utilizing speech modality as a trigger. 

For multimodal fusion, the use of deep neural networks has been explored  previously \cite{ngiam2011multimodal, wu2014survey, meng2020survey}.
Gomaa et al. study gaze and pointing modality for the driver's behavior while pointing to outside objects \cite{gomaa2020studying}. They further proposed various machine learning methods, including deep neural networks for a personalized fusion to enhance the predictions \cite{gomaa2021ml}. Aftab et al. demonstrate how multiple inputs, namely, head pose, gaze and finger pointing gesture, enhance the predictions of the driver's pointing direction for object selection inside the vehicle \cite{aftab2020you} and what limitations arise when pointing to objects outside the vehicle \cite{aftab2021multimodal}.

In our approach, we use a model-level fusion approach for selection of a wide range of objects that may be situated inside the vehicle or outside the vehicle. As head pose and gaze direction are directly related in identifying visual behavior \cite{ji2002real, mukherjee2015deep}, we use these two modalities along with finger pointing for two tasks: i) to identify whether the object to be selected lies inside or outside the vehicle, and ii) to precisely predict the pointing direction in either case. Each of the three modalities is processed as equal input, and the network learns from the training data.

In summary, object selection inside the vehicle and driver queries to outside-vehicle objects have been rigorously studied. However, to our knowledge, no study simultaneously deals with objects both inside and outside the vehicle. We merge concepts from previous research work to perform a comparison of the two above mentioned types of pointing. Our work differs from past work in that we deal with both in-vehicle objects as well as outside-vehicle objects. We compare and demonstrate modality specific limitations in both types of pointing and learn to distinguish between the two types. While most studies use simulators, we perform our experiments in a real car within authentic environment which both gives drivers a relatively realistic impression and helps us achieve more genuine and applicable results. Furthermore, unlike some of the related work, we use non contact sensors for tracking eye and gestures that allows users to behave more naturally. 

\section{Experiment Design}
For the application of deictic referencing with finger pointing and gaze, we used a real vehicle for data collection. For simplicity and ease of data collection, the vehicle was kept stationary at different locations on the road. Consequently, during all the pointing events at various objects inside and outside the vehicle, the primary focus of the driver was not on driving the vehicle, as would be the case in the self-driving cars in the future. Various non-contact and unobtrusive sensors were used to measure the drivers' gestural, head and eye movements. 

\subsection{Apparatus} \label{sec:apparatus}
We set the apparatus up in the same way as by Aftab et al. \cite{aftab2020you, aftab2021multimodal}. Two types of camera systems were used: 1) Gesture Camera System (GCS) and 2) Visual Camera System (VCS). The two camera systems were carefully chosen and consist of sensors placed in positions which are being used in production vehicles, e.g., BMW 7-series and BMW iX offer a camera fitted behind steering wheel to analyze the driver’s face as well as another camera at the car ceiling for gesture controls.

The GCS, which was mounted on the car ceiling next to the rear-view mirror, consisted of a Time-of-Flight (ToF) 3D camera with a QVGA resolution (320 $\times$ 240). It tracked both of the driver's hands for gestures and detected "one finger pointing gesture" from either or both hands, providing the 3D position of the finger tip as well as the direction of the pointing gesture as a 3D normalized vector. The direction vector was calculated from the 3D position of the finger base to the 3D position of the finger tip, and normalized to have a unit norm.  

The VCS, installed behind the steering wheel, captured the driver's head and eye movements. It provided the 3D position of the head center and the eyes along with the head orientation (as euler angles) and gaze direction as a 3D vector with unit norm.

Apart from these two camera systems, four additional cameras were placed inside the car, two of which recorded the driver's actions while the other two recorded the environment. These four cameras were used to analyze the events visually.

For speech, the Wizard-of-Oz (WoZ) method was used to note the timestamp of the speech command used with the pointing gesture. A secondary person (acting as a wizard) noted the instant (hereafter called the WoZ timestamp) when the driver made the referencing gesture and said, "what is \textit{that}?" with the help of a push button. This timestamp was used to identify the approximate time when the gesture took place.

\subsection{Feature Extraction} \label{sec:features}
From the two camera systems, GCS and VCS, we extracted the finger pose, eye pose and head pose, where pose constitutes both position and direction of the modalities. In total, we have six features. These are explained as follows:
\begin{itemize}
    \item Finger pose: the 3D position of the finger tip and (normalized) direction vector of the finger pointing gesture in the 3D vector space.
    \item Eye pose: the 3D position of the point between the two eyes and (normalized) direction vector of the eye gaze in the 3D vector space.
    \item Head pose: the 3D position of the center of the head and the Euler angles (as yaw, pitch and roll) of the head orientation.
\end{itemize}

\subsubsection{Pre-processing Data}
For each pointing event, we extracted a time interval of 0.8 seconds such that it included the WoZ timestamp (denoting the time of the speech command) within it. The duration of the time interval was based on the observation by Ru\"melin et al.\cite{rumelin2013free} for comfortable pointing time. This interval amounted to 36 frames (at 45 frames per second), forming a short temporal sequence. We used the whole temporal sequence for the model training explained later in Section \ref{sec:fusion}.

During the data collection, some of the referencing events contained occlusion in one or more modalities, which resulted in some frames with a few missing features. The occlusion in eye pose or head pose mainly occurred when the arm, that was used to point, was held in front of the face (as in Figure \ref{fig:teaser}) or when the head was turned to the far sides preventing the tracking of the eyes. In a few cases while pointing, the participants extended their arms beyond the field-of-view of the gesture camera, especially when pointing with the left hand to the left side, resulting in missing features of the finger modality. In order to fill in the missing features, we used linear interpolation from the two nearest neighbouring frames. 

\subsubsection{Axes Translation}

\begin{figure}[t]
\centering
      \includegraphics[trim=0 0 0 0,clip, width=0.478\textwidth]{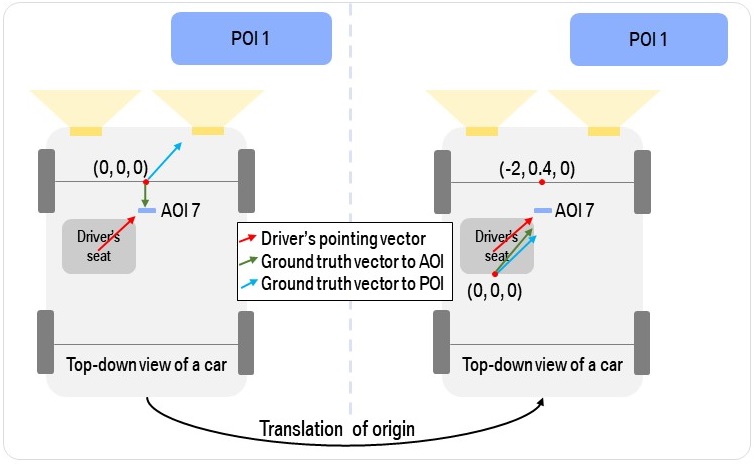}
\caption{Translation of origin from the center of the car's front axle to driver's seat}\label{fig:origin}
\end{figure}

\begin{figure}
\centering
  \includegraphics[trim=0 0 0 0,clip, width=0.478\textwidth]{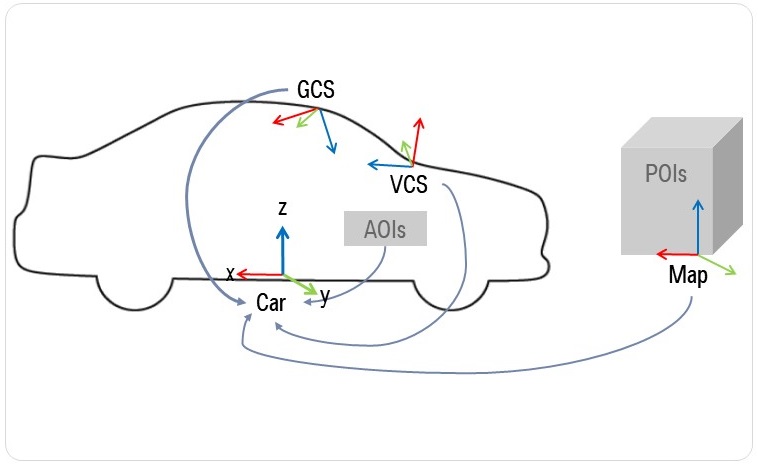}
\caption{Camera systems, AOIs, and map of POIs are converted to car coordinate system}\label{fig:car_coordinates}
\end{figure}

We followed the ISO 8855 \cite{iso2011road} standard for the car coordinate system with the exception of the origin's position (see Figure \ref{fig:origin}). We did not choose the origin at the center of the front axle because the inside objects lie before the front axle of the car and the outside objects lie ahead of car's axle. Consequently, having the origin at the center of the front axle, the ground truth vector, which is calculated from the origin to the center of the AOI or POI (defined in Section \ref{sec:gt}), seemingly becomes in opposite directions for inside and outside objects. For example, consider an object inside the vehicle, AOI 7, and an object outside the vehicle, POI 1. Despite the two objects lying in almost the same horizontal direction from the driver's perspective (i.e., approximately similar yaw angles for both) as shown in Figure \ref{fig:origin}, the ground truth vectors have a difference of about 130°. Therefore, for a fair comparison of pointing to inside and outside objects, and to have a comparable ground truth direction, we translate the origin by $[x=2\text{m}, y=-0.4\text{m}, z=0\text{m}]$, such that the origin resides at the approximated center point behind the driver's seat. Throughout the paper, this is kept fixed for all experiments for consistency. With this translation the ground truth vectors have a similar yaw direction for both objects (see Figure \ref{fig:origin}, right). Consequently, all features from both cameras were transformed to the car coordinate system, with the origin behind the driver's seat (see Figure 3).

\subsection{Experiment Types}
The experiments were divided into two types, the cockpit use case and the environment use case. For both of these types, the apparatus and the vehicle were kept the same, and vehicle was kept stationary. In both use cases, the participants were asked to point naturally to the pre-selected objects, and say "what is \textit{that}?". They were free to choose either hand for pointing. Some objects were larger than others and drivers could choose to point to any visible part of the  surface area. The difference between the two use cases lay in the chosen objects. Consequently, the pointing directions differed as well as the angular width and angular height of objects. The objects in both use cases were chosen such that they were in front of the driver, including the far right as well as far left sides with respect to the driver, in order to have a sufficiently large variance of direction angles in both cases.

\subsubsection{Cockpit Use Case}

\begin{figure}[h]
\begin{subfigure}{.478\textwidth}
  \centering
  \includegraphics[trim=0 40 0 10,clip, width=\textwidth]{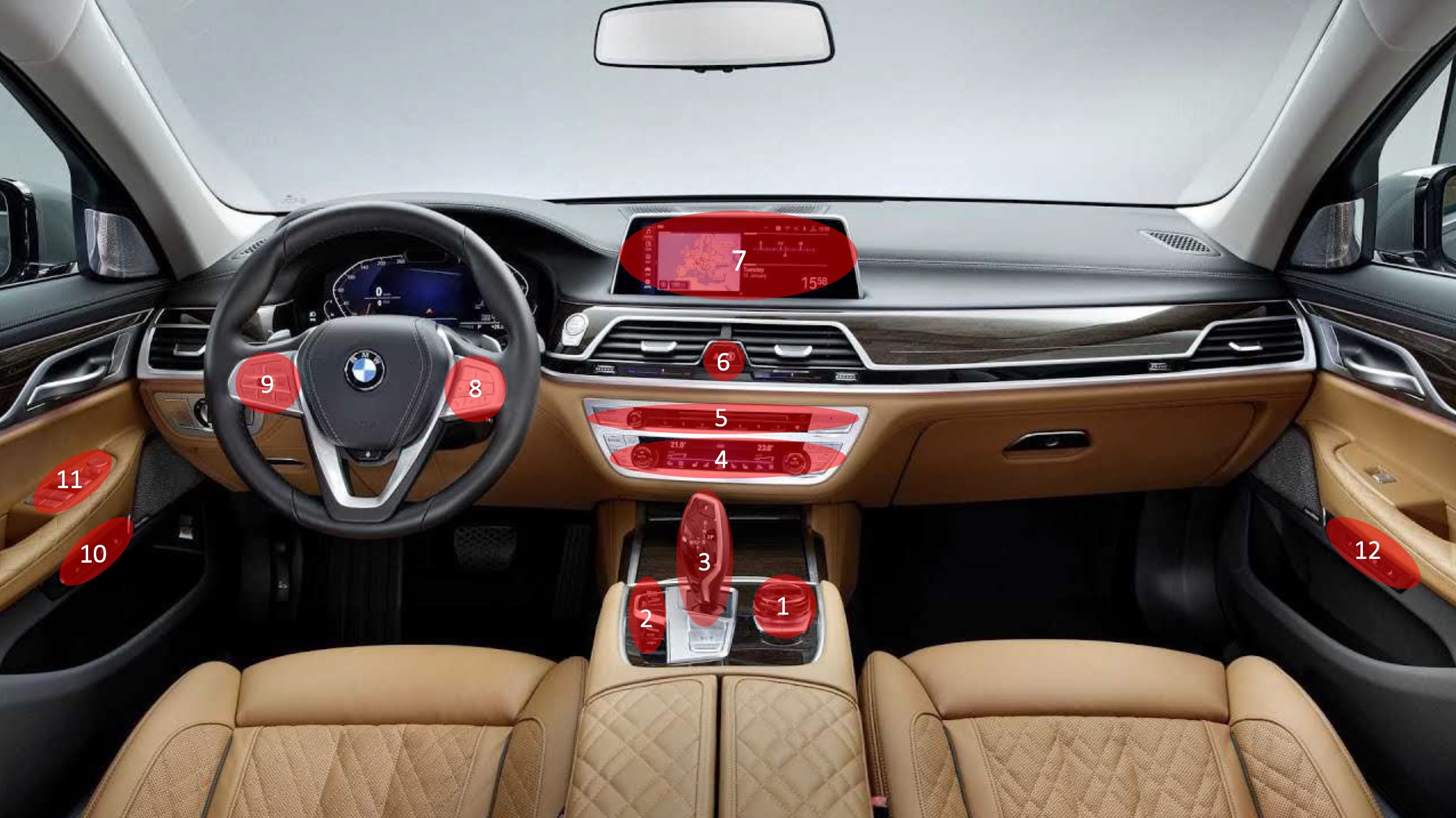}
  \caption{AOIs shown with red highlighted areas}
  \label{fig:AOIs_car}
\end{subfigure}
\begin{subfigure}{.478\textwidth}
  \centering
  \includegraphics[trim=0 0 0 0,clip, width=\textwidth]{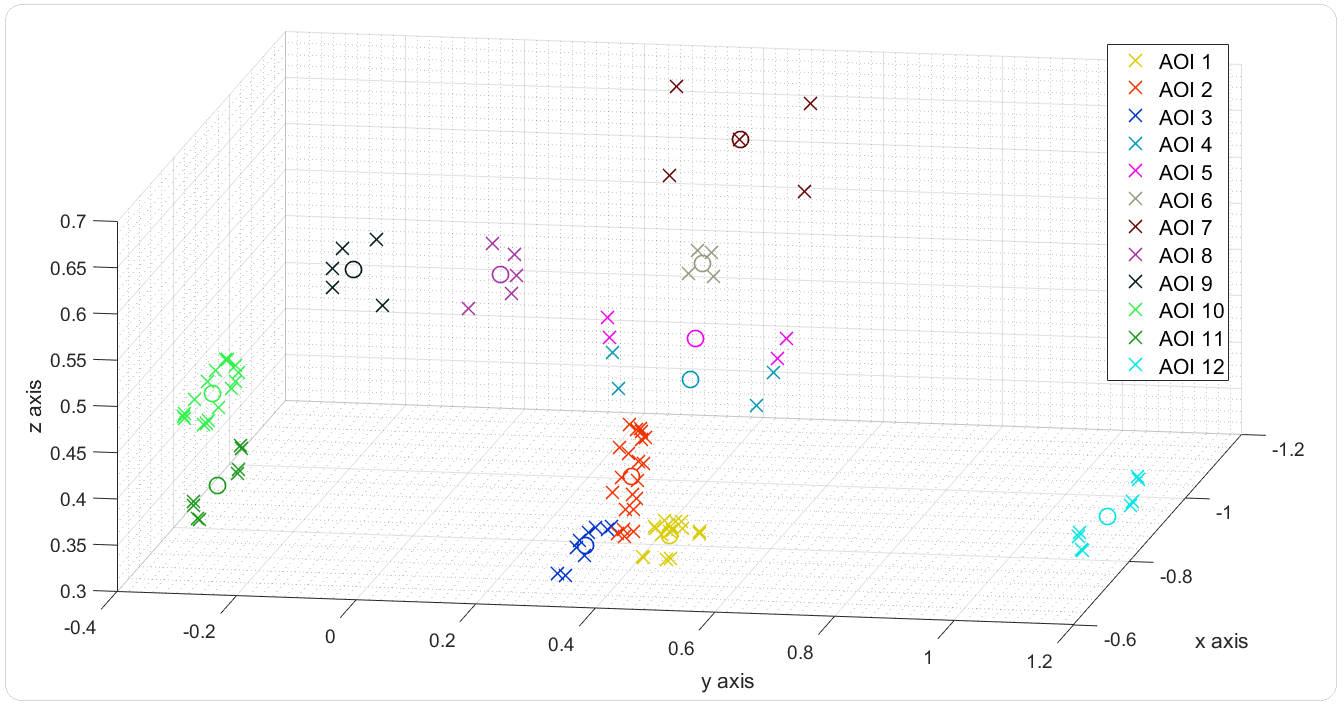}
  \caption{Measured corner points of AOIs}
  \label{fig:AOIs_scatter}
\end{subfigure}
  \caption{The 12 selected AOIs in the cockpit of a car (left), and the scatter plot of the measured corner points of the AOIs w.r.t the car origin (right) \cite{aftab2020you}. }
  \label{fig:AOI}
\end{figure}

 In the first type of experiment, 12 distinct areas inside the vehicle were chosen, called the Areas-of-Interest (AOIs). These were different control elements of the car that the driver could potentially reference for touch-free control, shown in Figure \ref{fig:AOIs_car} illustrated by red circles. Figure \ref{fig:AOIs_scatter} shows the measured points of the corners or vertices of the AOIs with crosses 'x', and the mean point of each AOI with a circle 'o'. These define the areas where the users in this first type of experiments should point to. Consequently, the AOIs have different (but fixed) sizes at chosen distances and locations as shown in Figure \ref{fig:AOIs_scatter}.
 
 Referencing of AOIs was independent of car position. All participants were asked to point to the given 12 AOIs for 0 times, not necessarily in the same sequence. However, not all samples could be correctly recorded and therefore, had to be discarded due to technical issues with the setup. In total, we had 2514 samples that were used for training and testing for the cockpit use case.

\subsubsection{Environment Use Case}

\begin{figure*}[t]
  \includegraphics[trim=0 0 0 0,clip, width=\textwidth]{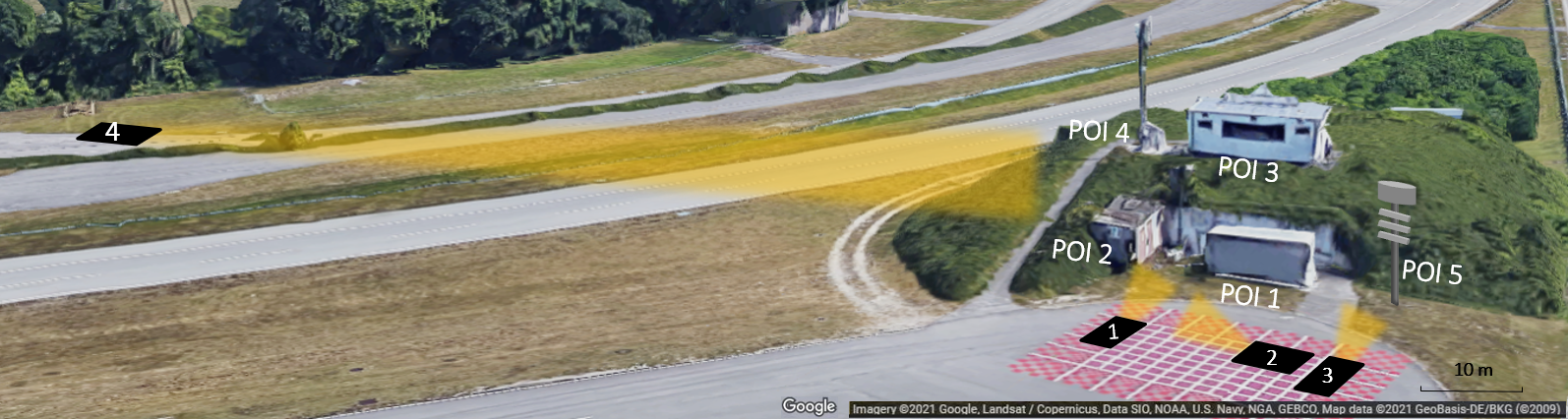}
  \caption{Map of the chosen 5 POIs along with the 4 car poses shown by black rectangles \cite{aftab2021multimodal}.}
  \label{fig:map}
\end{figure*}

In the second type of experiment, 5 different landmarks situated in front of the vehicle were chosen. The landmarks were buildings and antennas. These are referred to as Points-of-Interest (POIs). Referencing of POIs was conducted in 4 different car poses\footnote{GPS coordinates 1: 48.220446N 11.724796E, 2: 48.220363N 11.724800E, 3: 48.220333N 11.724782E, 4: 48.221293N 11.724942E}, where pose constitutes both position and orientation, to add a large variety of pointing directions. The coordinates of the car poses and the POIs were manually measured with a laser sensor, Leica Multistation\footnote{{https://leica-geosystems.com/en-us/products/total-stations/multistation}}, and converted from geodetic coordinates to Cartesian coordinates with origin at the driver's seat. The POIs and car poses in the environment use case are shown in Figure \ref{fig:map}. It can be seen that the outside objects or POIs in this stationary use case also have different but fixed sizes, each with fixed distance and location w.r.t. the vehicle.

In the fourth car pose, only 3 out of the 5 POIs were visible to the driver. Consequently, the 4 car poses and 5 POIs provided 18 different pointing directions in front of the driver as well as on the right and left sides of the driver (see car position '2' and '3' in Figure \ref{fig:map}). 

Similar to the cockpit use case, the participants were asked to point to each POI from each of the four different car poses. The car pose and POI were repeatedly changed so that users did not get accustomed to the next POI to be referenced. We collected 6590 samples for the environment use case. The reason for the relatively larger data collection for the environment case is two fold: i) there were more pointing directions, and ii) we needed a larger variance in data to get adequate results and to obtain a robust model for environment use case. This was because of the relatively larger pointing errors by the users in the environment use case as compared to the cockpit use case  - as can be seen in Section \ref{sec:analysis}, Figures \ref{fig:measurements_yaw} and \ref{fig:measurements_pitch}.

\subsection{Participants and Data Collection}
For our experiments, thirty participants took part in at least one of the two experiments. However, for the sake of fair comparison, we only considered those 11 participants which took part in both experiments. The participants ranged from 20 years old to 40 years old, with a mean age of 28.7 and a standard deviation of 5.7. Two of the eleven participants were females. Three participants wore glasses and one wore contact lenses, while the rest did not wear any glasses or lenses. Only one participant was left handed. However, the hand used for gesture by the right handed users was not always the right hand. In the cockpit use case, 23\% of the events were performed with left hand. In the environment use case, about 12\% were carried out with the left hand. It is important to mention here that due to a few administrative and technical reasons, the number of samples per driver are not perfectly balanced. Furthermore, we collected more samples for the environment use case than the cockpit use case. 

\subsection{Dataset Split} \label{sec:data_split}
We split the dataset into three sets: training set, validation set and test set. The division of the sets was participant based. This means that no reference sample from participants in the training set appeared in either the validation or the test set, and vice versa. This ensures real-world validity. For generalization, we used a leave-one-out cross validation to evaluate our models. The leave-one-out split resulted in 11  splits of the dataset that were used for testing as we had 11 participants. Weighted average is used to calculate the final metrics. In this, the entire dataset is covered in the test set. For each test split, a different participant (not present in the training set), was used for validation. Consequently, we had 11 splits of the training set as well, each with a different subset of the participants. 
  
\subsection{Analysis of Modality Measurement} \label{sec:analysis}
\subsubsection{Preciseness of Measured Modalities}

\begin{figure}[t]
\begin{subfigure}{.478\textwidth}
 \centering
  \includegraphics[trim=0 0 0 0,clip, width=\textwidth]{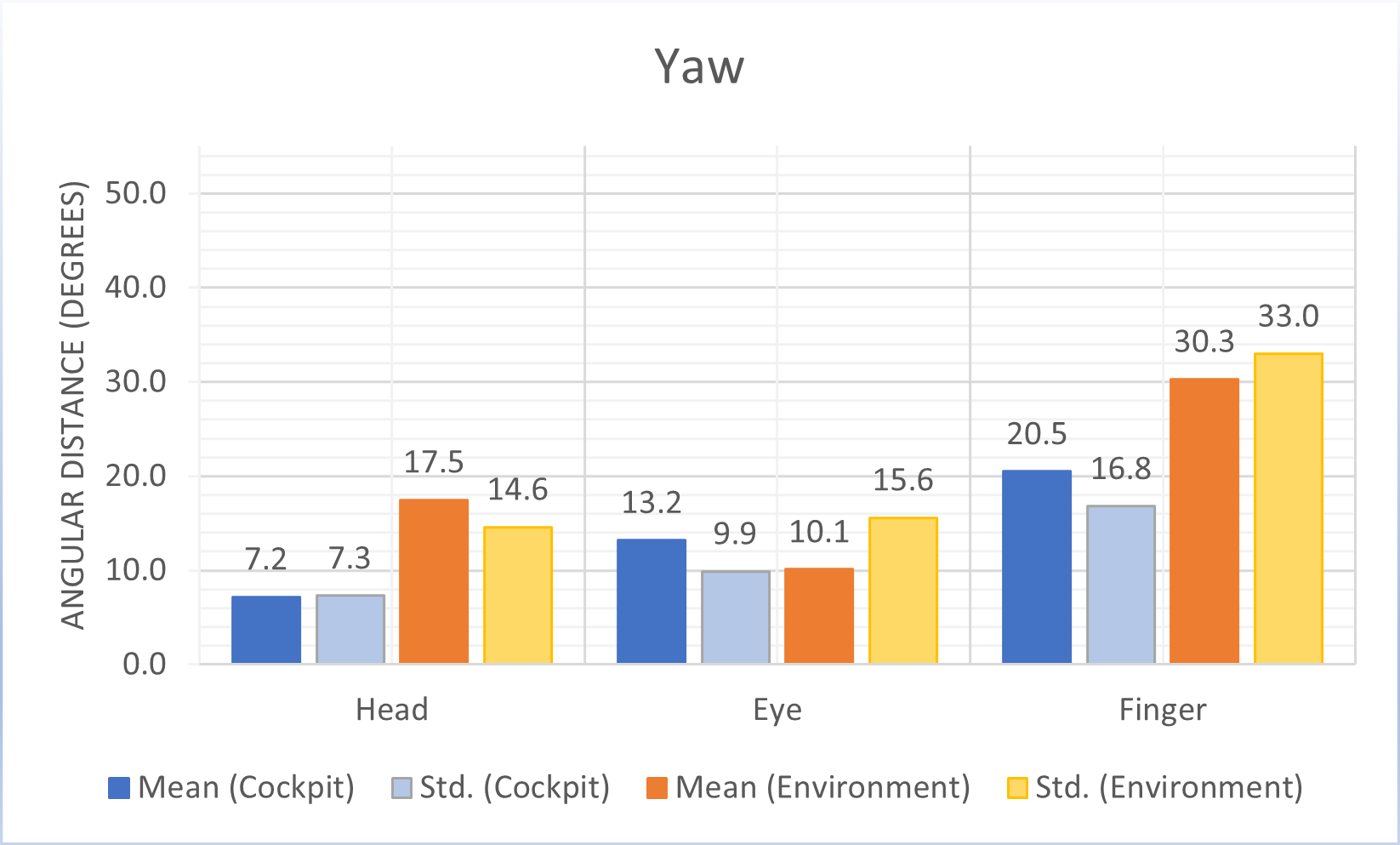}
  \caption{Mean and Std. in yaw (horizontal) angles}
  \label{fig:measurements_yaw}
\end{subfigure}
\begin{subfigure}{.478\textwidth}
      \centering
  \includegraphics[trim=0 0 0 0,clip, width=\textwidth]{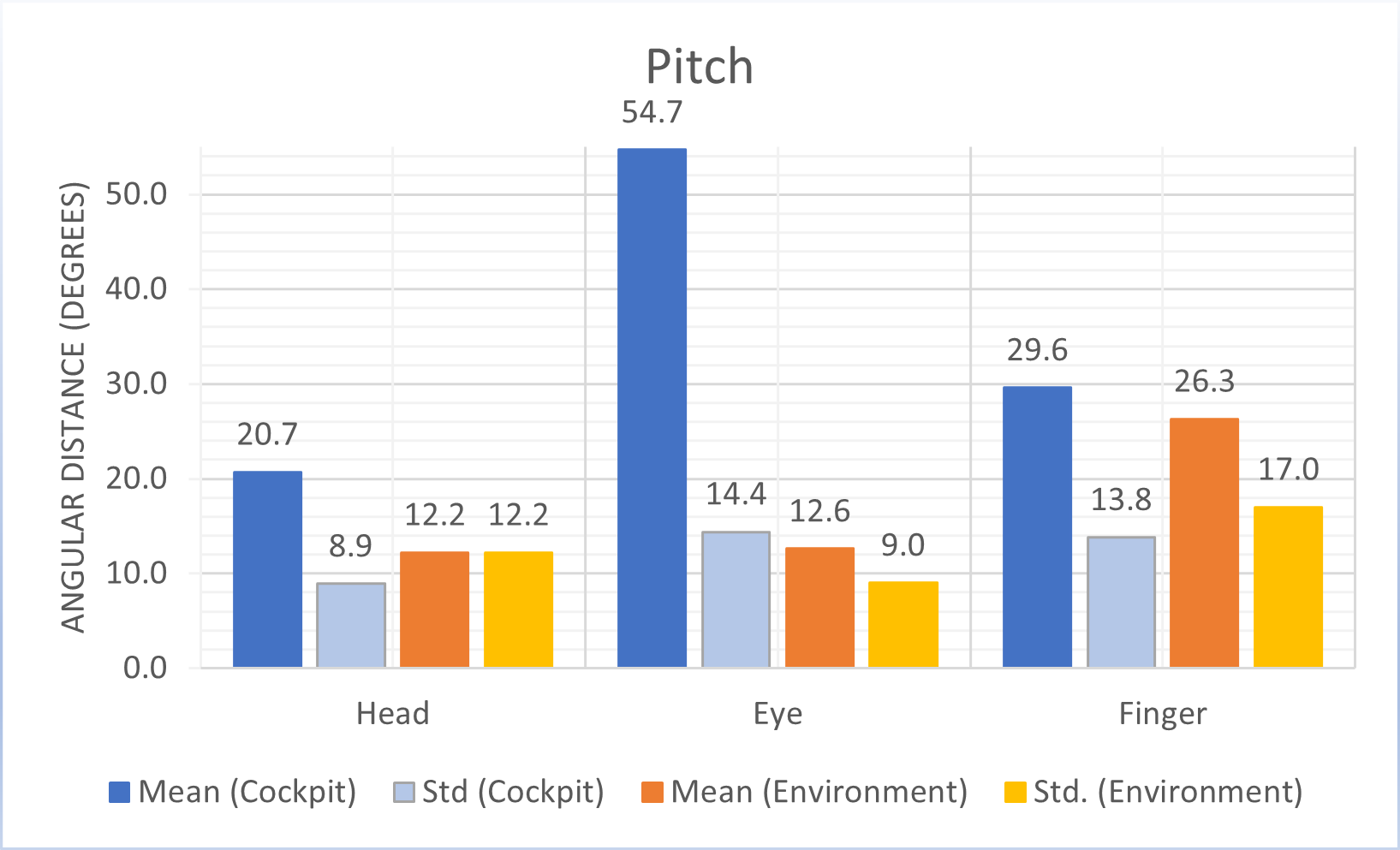}
    \caption{Mean and Std. in pitch (vertical) angles}
      \label{fig:measurements_pitch}
\end{subfigure}  
  \caption{Mean and standard deviation (Std.) of the angular distance between measured direction of the modalities and the ground truth direction at the WoZ timestamp instant.}
  \label{fig:measurements}
  \end{figure}

We started by analyzing the quality of data collection. To this end, for each modality we calculated the mean and standard deviation of the angular distance between the measured direction and the ground truth direction at the instant when the WoZ button is pressed (i.e., at the WoZ timestamp). This instance lies in the middle of the pointing gestures in almost all events. Only for simplifying the analysis of the 3D vectors, all coordinates were converted from Cartesian ($x,y,z$) to spherical coordinates ($r, \theta, \phi$). The mean and standard deviation of the modalities for the two use cases are illustrated in Figures \ref{fig:measurements_yaw} and \ref{fig:measurements_pitch} for the yaw and pitch directions, respectively.

It can be seen that in the environment use case, the measured direction angles w.r.t. ground truth for the finger modality (30.3° in yaw and 26.3° in pitch) are significantly larger than the other two modalities in both yaw and pitch. Furthermore, the standard deviations of the finger direction w.r.t. ground truth (33.0° in yaw and 17.0° in pitch) are also large than the standard deviation of eye direction and head direction. This might be caused by the relatively low availability of the finger modality in the environment use case (discussed in more details in sections \ref{sec:results_environment} and \ref{sec:cross_dataset}). 

In the cockpit use case, we observe a relatively large angular distance in the pitch angles for all three modalities. The eye direction has the largest angular distance in the pitch angles (54.7°) in comparison with the other two modalities. This larger angular deviation of eye direction stems from the position of the origin lying below the eye position. As the ground truth vector is calculated from origin, the pitch direction of ground truth for all AOIs and POIs would be upwards, whereas the eye direction for the majority of the AOIs would be downwards. Therefore, relatively high values were measured for the eye direction as well as the head direction towards AOIs in the cockpit use case. However, the pitch angles of the head (20.7°) exhibit a much smaller offset as compared to the pitch of the eye direction. This indicates that, on average, the head direction was not entirely turned towards the AOI. 

\subsubsection{Distribution of Direction Angles}
  
\begin{figure}[t]
\begin{subfigure}{.236\textwidth}
    %  \centering
  \includegraphics[trim=0 0 0 0,clip, width=\textwidth]{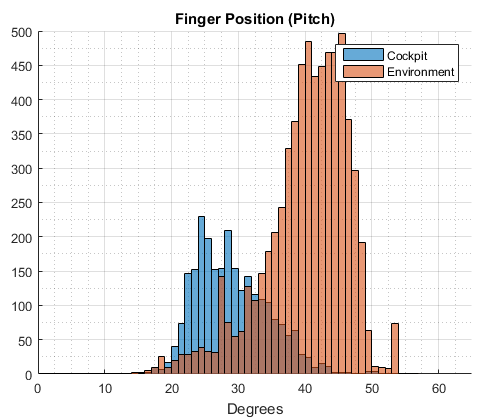}
    \caption{Finger position (pitch)}
    \label{fig:fngr_pos}
\end{subfigure}  
\begin{subfigure}{.236\textwidth}
    %  \centering
  \includegraphics[trim=0 0 0 0,clip, width=\textwidth]{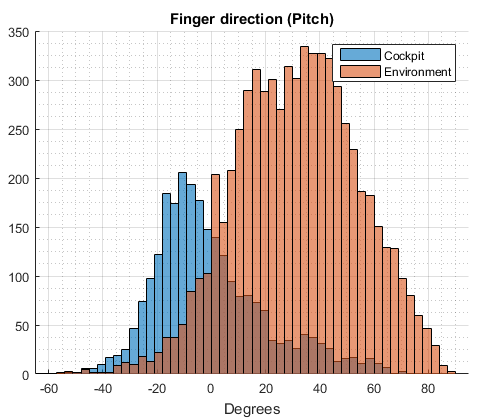}
    \caption{Finger direction (pitch)}
        \label{fig:fngr_dir}
\end{subfigure}  
\begin{subfigure}{.236\textwidth}
    %  \centering
  \includegraphics[trim=0 0 0 0,clip, width=\textwidth]{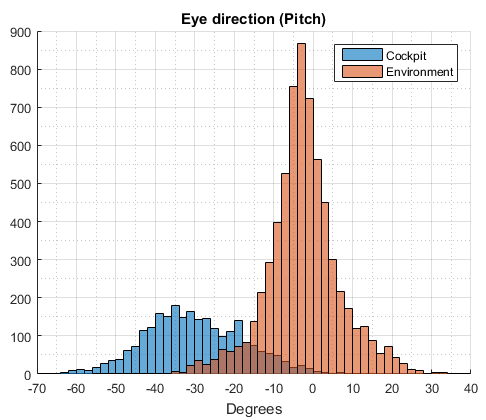}
    \caption{Eye direction (pitch)}
        \label{fig:eye_dir}
\end{subfigure}  
\begin{subfigure}{.236\textwidth}
    %  \centering
  \includegraphics[trim=0 0 0 0,clip, width=\textwidth]{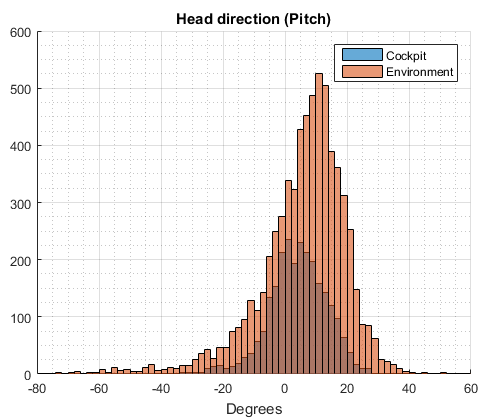}
    \caption{Head  direction (pitch)}
        \label{fig:head_dir}
\end{subfigure}  
  \caption{Distribution of the pitch (vertical) angles of (a) finger position, (b) finger direction, (c) eye direction and (d) head direction.}
  \label{fig:hists}
  \end{figure}
  
The distribution of the direction of the eye, head and finger pointing is vital to understand the contrast between pointing inside the vehicle and outside. As the control elements mainly lie below the windscreen, the referencing direction for the AOIs would mostly be downwards while for the POIs, it would be slightly upwards or parallel to the road. This trend is observed in our data collection, which is shown by distribution of the modalities' (pitch) direction in Figure \ref{fig:hists}. From the distributions of the two cases, we see a clear separation in the eye direction in Figure \ref{fig:eye_dir}. The finger direction has overlapping distributions (see Figure \ref{fig:fngr_dir}), whereas the head direction has no separation in pitch directions at all (see Figure \ref{fig:head_dir}). Interestingly, the finger position also plays a distinctive role in differentiating between the types of referencing objects (see Figure \ref{fig:fngr_pos}). For all drivers, the pitch angles of the finger tip position w.r.t. the origin at the center point behind driver's seat for the cockpit use case has a mean and standard deviation of 29.0° and 5.4°, respectively (see Figure \ref{fig:fngr_dir}). Whereas, for the environment use case, considering all drivers, the pitch of the finger tip position has a mean of 40.0° and a standard deviation of 6.8°.  
  
\section{Multimodal Fusion Models}
\begin{figure*}[t]
     \centering
  \includegraphics[trim=0 0 0 0,clip, width=\textwidth]{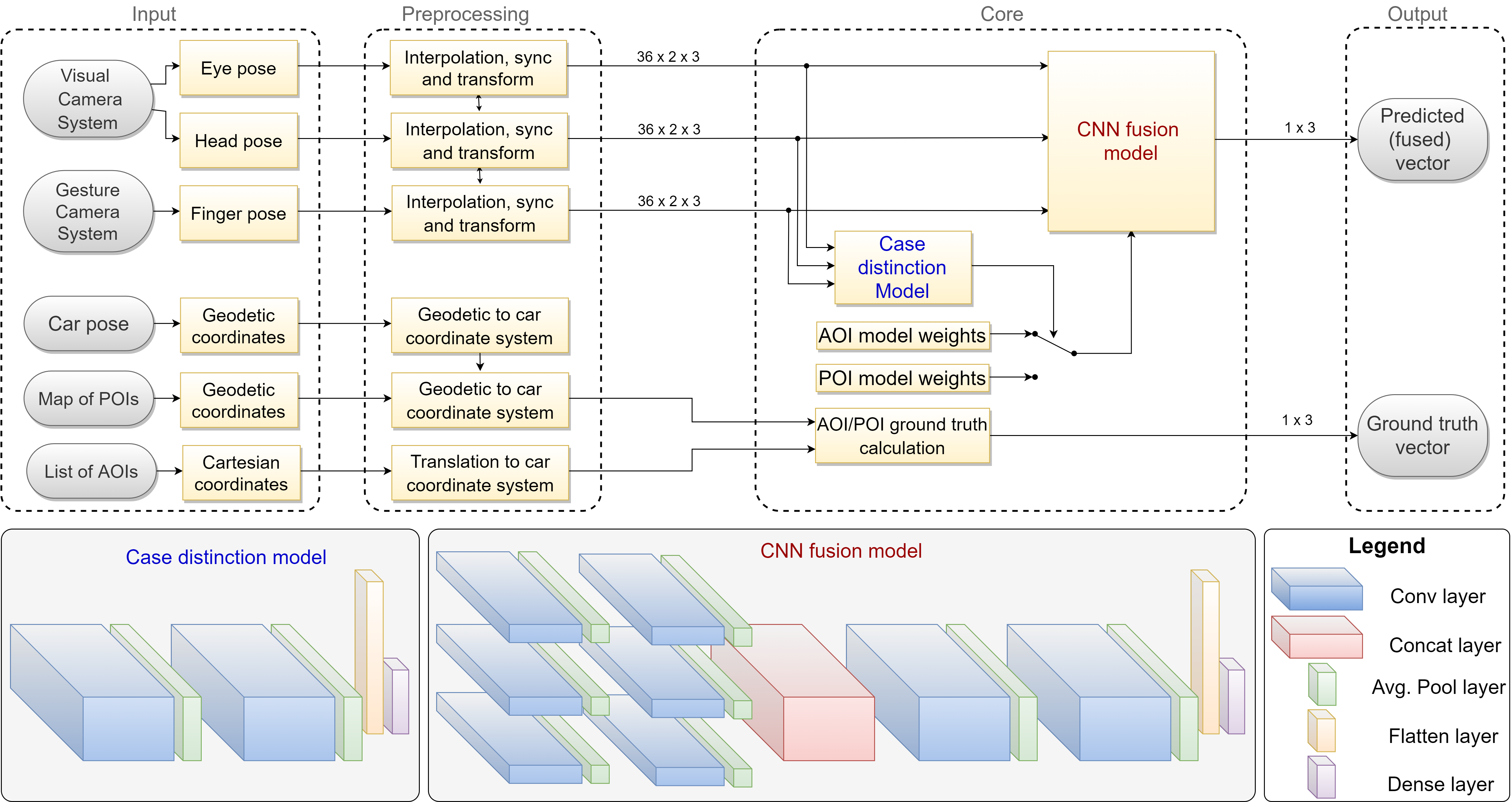}
      \centering
  \caption{The overall architecture of our approach.}
  \label{fig:methodology}
  \end{figure*}

We propose a two-step approach, illustrated in the overall architecture in Figure \ref{fig:methodology}, for the recognition of use case type and for the fusion of modalities.
  
\subsection{Case Distinction Model}
Firstly, taking the six features as input, for early fusion, we use a shallow Convolutional Neural Network (CNN) model (consisting of 2 convolutional layers and 1 dense layer) with a binary cross-entropy loss to predict whether the driver referenced an object inside or outside the car. Each of the two convolutional layers contains 64 kernels with a size of $3 \times 3$. The shallow model for use case distinction is trained independently from the fusion model (which is discussed in Section \ref{sec:fusion}). The model consisted of approximately 21,000 trainable parameters.

The output of the model is used to load the appropriate weights for the fusion model, which is  trained separately for the two use cases. If the referenced object is determined to be an AOI, the weights of the cockpit use case are applied to the fusion model, and if the referenced object is determined to be a POI, the weights of the environment use case are applied to the fusion model. 
  
\subsection{Fusion Model} \label{sec:fusion}
For each referencing event, we aim to predict the direction angle towards an AOI or POI w.r.t the ground truth. Therefore, the problem we deal with is a regression problem as the output is continuous. We adopt a model-level fusion method \cite{chen2016multi} for integration of the set of pre-processed features mentioned in Section \ref{sec:features}. The model-level fusion also has the tendency to implicitly learn the temporal relations between modalities \cite{wu2014survey}. A deep CNN is designed having linearly regressed output, which is trained on the collected data from the 11 participants. The motivation behind the choice of the deep CNN model is to cover the large number of behavioral cases exhibited by the participants which a rule-based or a simple linear regression model may not be able to cover. As the input features form a temporal sequence, the convolution block operates in the temporal dimension as well as the feature dimension. The overall architecture is shown in Figure \ref{fig:methodology}. 

\subsubsection{Model Description}
The input, $x$, to the CNN model is a batch of size $b$, such that $x \in \mathbb{R}^{b \times t \times f \times d}$, where  $t=36$ is the number of temporal (consecutive) frames that form the sequence, $f=6$ is the number of features (2 for each modality) and $d=3$ is the number of dimensions in each feature (cartesian coordinates). The CNN model consists of two convolutional layers, applied on each modality (eye, head and finger) separately, each with a kernel size of $2 \times 2$ and Rectified Linear Unit (ReLU) activiation function, followed by an average pooling layer of size $2 \times 1$. 

The convolutional feature maps are then concatenated and two more convolutional layers of kernel size $3 \times 3$ with ReLU activiation functions are applied. The convolutions in these layers share the information from each modality. The number of kernels for each convolution layer is selected to be 128. A flatten layer is used to vectorize the feature maps before finally applying a fully connected layer, which uses linear activation to provide the linearly regressed fused 3D direction vector, $y \in \mathbb{R}^{b \times d}$. The CNN model has approximately 0.5M parameters when all three modalities were used as inputs, approximately 0.43M parameters when two modalities were used and about 0.36M parameters when only one modality was used.

\subsubsection{Ground Truth} \label{sec:gt}
In order to calcuate the ground truth in the cockpit use case, the measured points of the AOIs (shown in Figure \ref{fig:AOIs_scatter}) were translated to the car coordinate system. In the environment use case, the GPS coordinates of the eight corners of each POI were measured in WGS84 (World Geodetic System) standard. The geodetic coordinates of the POI corners were first converted to Cartesian Earth-Centered, Earth-Fixed (ECEF) and then, to the car coordinate system using an affine transformation with the rotation and translation matrices calculated from the car pose.
We, then, defined ground truth as the normalized 3D vector with unit norm calculated from origin (i.e. driver's seat) to the center of the AOI or POI, calculated by taking the mean of the measured corner points of AOI or POI, respectively.

\subsubsection{Loss function}
For the training of the network, Mean Angular Distance (MAD) between the output vector and the ground truth vector was used as the loss function, $\mathcal{L}$. In other words, the angle between the two vectors is minimized. Mathematically:
% \begin{equation}
%     \mathcal{L} = - \frac{1}{N} \sum^N_{i=1}  (\theta_i) 
%     = - \frac{1}{N} \sum^N_{i=1} \frac{\textbf{\^y}_i \ . \ \textbf{y}_i}{\|\textbf{\^y}_i \| \  \| \textbf{y} _i\|} 
%     \label{eqn}
% \end{equation}

\begin{align}
    \mathcal{L} = \text{MAD} &= \frac{1}{N} \sum^N_{i=1}  \theta_i \qquad \qquad \quad \qquad \in \ [0,\pi]     
    \label{eq:eq1} 
    \\
    &= \frac{1}{N} \sum^N_{i=1}  \text{arccos} \left( \frac{{\hat{\bm y}}_i \cdot \ {\bm y}_i}{\|{\hat{\bm y}}_i \| \  \| {\bm y} _i\|} \right)  
\end{align}

where $  \hat{\bm y}_i$ is the $i$-th 3-dimensional predicted vector, $\textbf{y}_i$ is the $i$-th 3-dimensional  ground truth vector, $\theta_i$ is the angle between the two 3D vectors, and $N$ is the total number of samples.

\subsection{Model Training}
For each modality or combination of modalities in both use cases, the shallow case distinction model and the CNN fusion model were trained separately. Each training was performed with the same parameters: a batch size of 32, Adam optimizer with a variable learning rate starting from 0.001, and 50 epochs. The model which minimized the validation loss was chosen to evaluate the test set. Finally, the weighted average of the test sets provided the performance metrics.

\subsection{Performance Metrics}

% \subsubsection{Measurement Precision}
% To evaluate the preciseness of measurements from the sensors for each modality, we use calculate the mean (M) and the standard deviation (SD) of the absolute angular distance of measured direction from the ground truth direction in the horizontal (yaw) and vertical (pitch) directions.

\subsubsection{Classification Accuracy}
To measure the performance of the case distinction model, we use a binary classification accuracy, as there are only two cases. The accuracy is the percentage of correctly identified cases, i.e., true positive rate for binary classification.   

\subsubsection{Mean Angular Distance (MAD) and Standard deviation of Angular Distance (Std.AD)}
We use two metrics to measure the performance of the fusion model in terms of precision. The MAD is used to evaluate the precision of the regression output from the model. It is defined as mean of the angular distance between the predicted and ground truth vectors. It is the same as the loss function used above and is mathematically shown in Eq. \ref{eq:eq1}. The smaller the angular distance between them, the more precise the prediction is. Therefore, lower MAD means better precision. We also calculate the standard deviation of the angular distance (Std.AD) to analyze the variation in the angular distances.  
  
\subsubsection{Hit Rate}

\begin{figure}[t]
     \centering
  \includegraphics[trim=0 0 0 0,clip, width=0.478\textwidth]{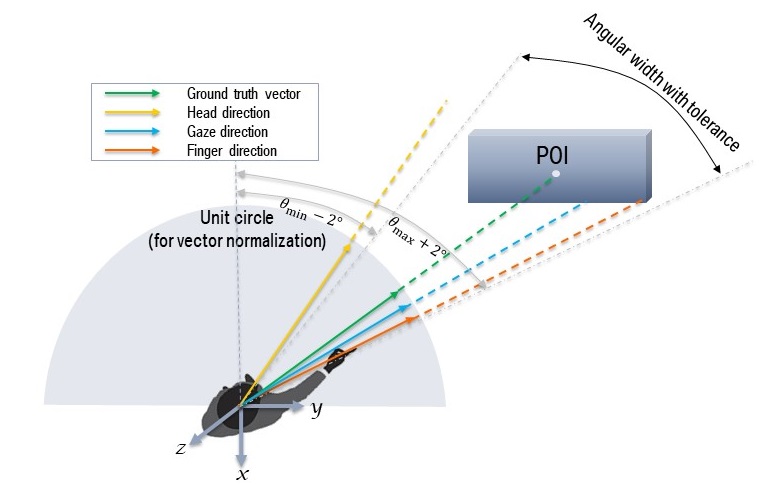}
      \centering
  \caption{Top-down view of user pointing to POI. Here, the direction vector of head is not counted as a hit as it is outside the angular range of the POI, while directions of gaze and finger are considered as hits.}
  \label{fig:hits}
  \end{figure}
  
For each pointing reference, a hit was considered if the direction angle of predicted (output) direction vector was within the range of angular width and angular height of the object (i.e AOI or POI) with a tolerance of 2° and 1°, respectively. This is illustrated in Figure \ref{fig:hits}. 
% For measured directions of each (input) modality, a hit was considered when the direction angle was within the angular range of POI/AOI with tolerance for atleast 200 millisecond. 
The hit rate was then calculated by dividing the sum of all hits by the total number of referencing events, shown in Eq. \ref{eq:eq2}.

\begin{align}
        hit\_rate& = \frac{\sum hit}{\sum    referencing\_events} \label{eq:eq2}  \\
        \text{and } \qquad hit &= 
        \begin{cases}
            1 \quad \text{if } \; (\theta_{min} - 2 < \theta < \theta_{max} + 2) \\ \text{ and } \ (\phi_{min} - 1 < \phi_{pitch} < \phi_{max} + 1)\\
            0 \quad \text{otherwise}
        \end{cases}
\end{align}
where $\theta$ and $\phi$ are the horizontal (hereafter called yaw) and vertical (hereafter called pitch) angles of the vectors (when converted to spherical coordinates), respectively, and $\theta_{min}$, $\theta_{max}$, $\phi_{min}$ and $\phi_{max}$ are the minimum and maximum angles of the angular width in yaw and pitch for the referenced object with respect to the car. 

It is important to note that the hit rate is not the accuracy of correct identification of the AOI or POI. The object (AOI or POI) can ,in some cases, still be correctly identified even though it was not hit, based on the closest cosine proximity. In the context of this paper, we do not compare the object identification accuracy as this depends upon the location and density of the objects, which are different in the cases we study.

\section{Experiments and Results} \label{exp}
In this section, we analyze and discuss the results obtained from our various experiments. For each experiment, we test our results using a weighted average from the 11-fold cross-validation as discussed before in Section \ref{sec:data_split}.

\subsection{Case Distinction Results}

\begin{figure*}[t]
     \centering
  \includegraphics[trim=0 0 0 0,clip, width=\textwidth]{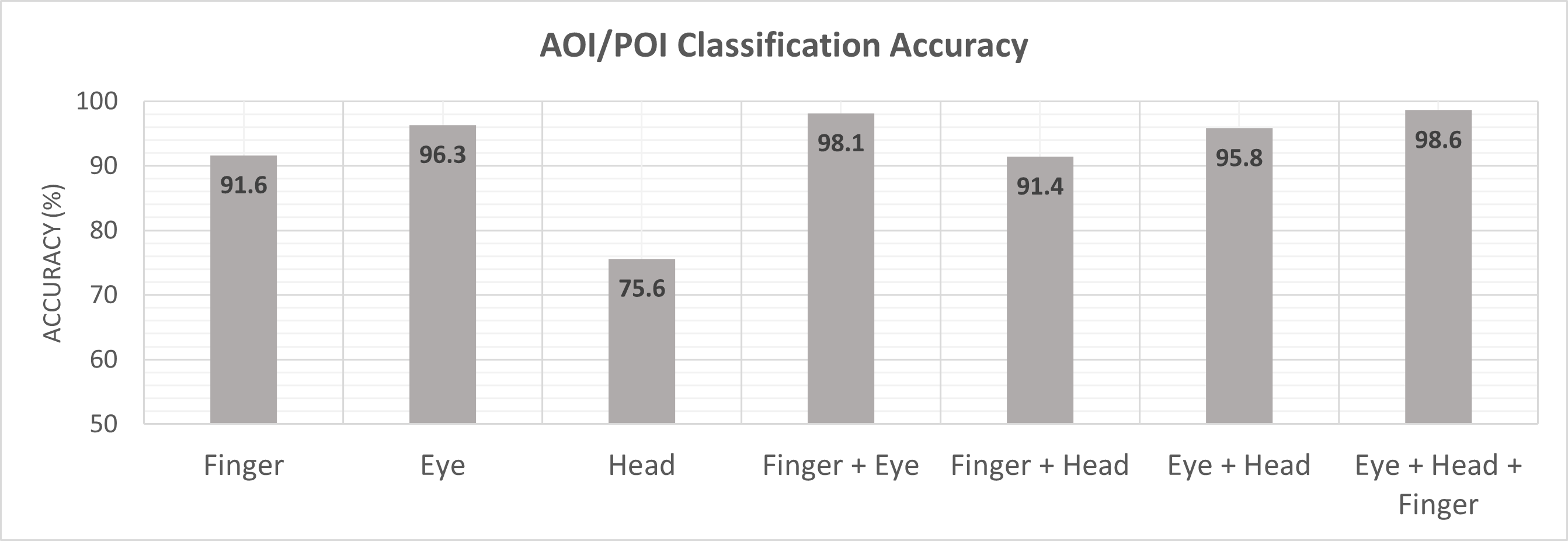}
      \centering
  \caption{Classification accuracies for use case distinction for all modalities}
  \label{fig:distinction_accuracy}
  \end{figure*}
   
As there is a considerable difference in the distributions of the pitch angles of the modalities, especially finger direction and position for the two use cases (shown in Figures \ref{fig:fngr_pos} and \ref{fig:fngr_dir}), we expected a simple CNN model to handle the classification task. While using all six features from the three modalities, an 11-fold accuracy for classification of 98.6\% was achieved (see Figure \ref{fig:distinction_accuracy}). We observed eye gaze as the dominant contributor with an accuracy of 96.3\%, which was further enhanced by the finger modality to 98.1\% possibly because of the different finger positions for inside and outside pointing. When using only finger pose as input to the model, a classification accuracy of 91.6\% was achieved.

\subsection{Ablation Study: Modality Specific and Fusion Results}

\begin{figure*}[t]
     \centering
  \includegraphics[trim=0 0 0 0,clip, width=\textwidth]{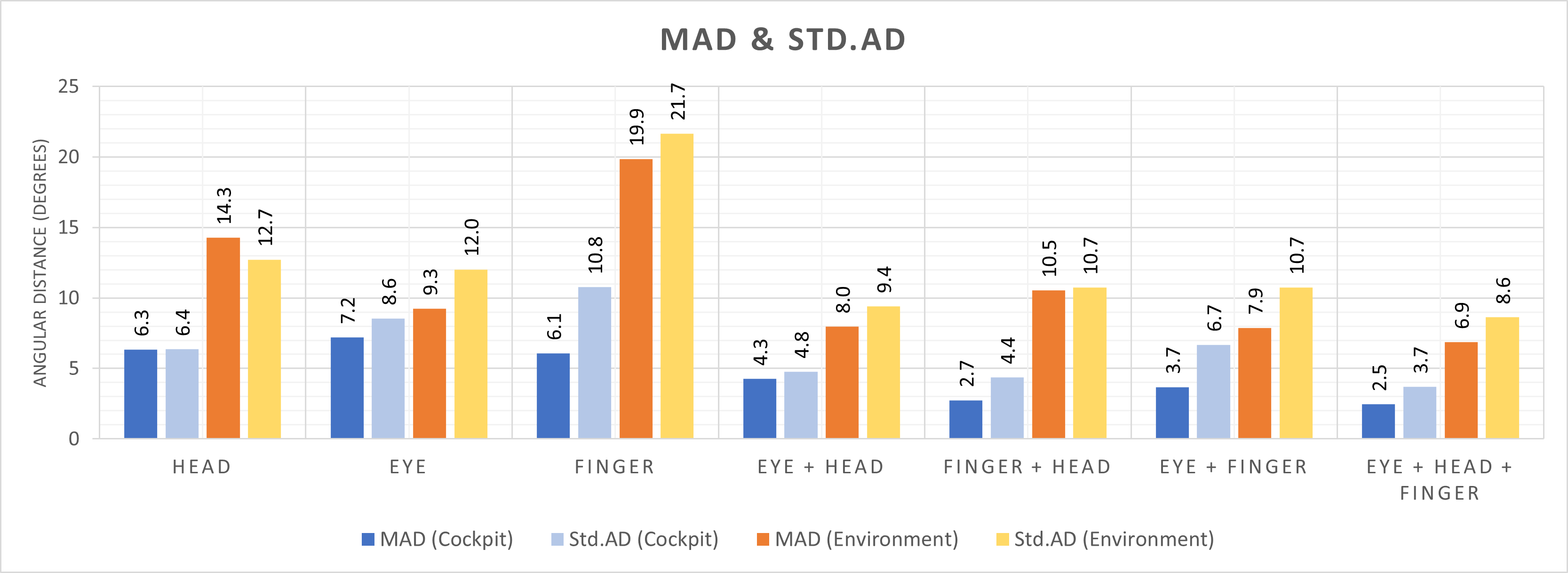}
  \caption{MAD and Std.AD of the resultant for different combinations of modalities in the cockpit and environment use cases}
  \label{fig:MAD}
  \end{figure*}

\begin{figure*}[t]
       \centering
  \includegraphics[trim=0 0 0 0,clip, width=\textwidth]{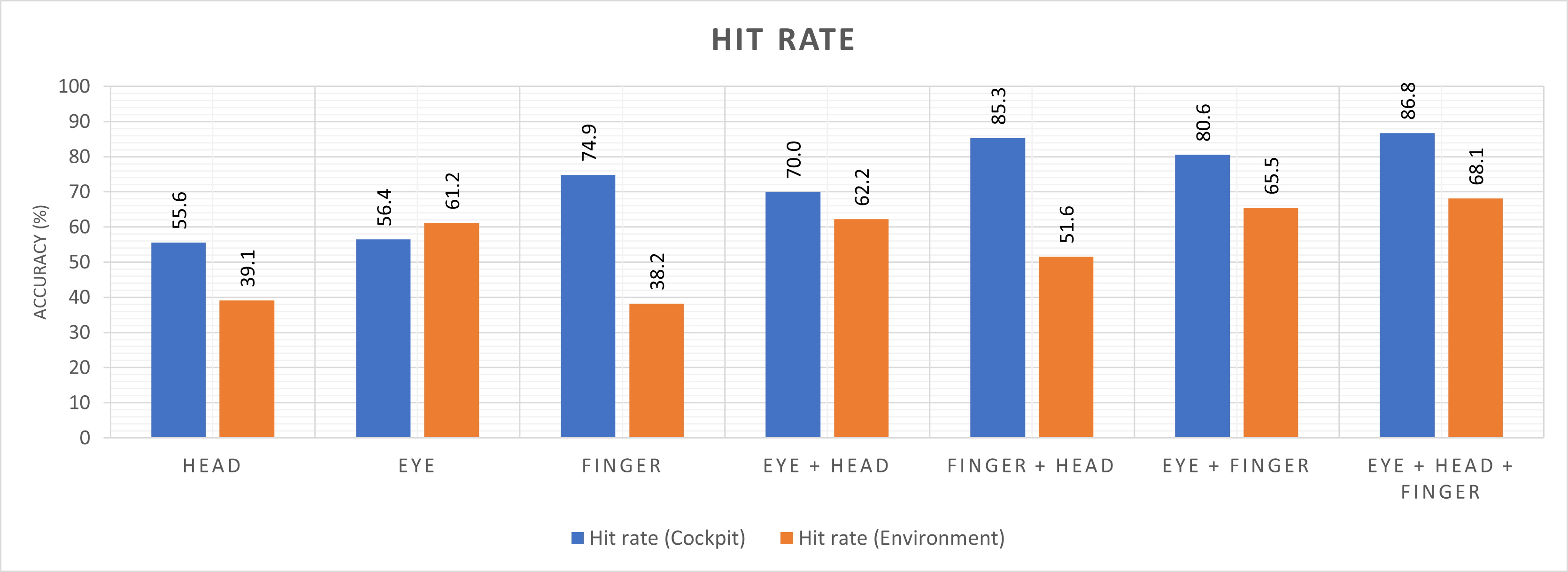}
      \centering
  \caption{Hit rates of the resultant for different combinations of modalities in the cockpit and environment use cases}
  \label{fig:hit_rate}
\end{figure*}

The ablation study helps to understand the effects of different components of the network. We use single modalities, a combination of two modalities and all three modalities simultaneously to train the fusion model and analyze the effects of adding modalities in both cockpit and environment cases. Figures \ref{fig:MAD} and \ref{fig:hit_rate} show a comparison of the results for all modality combinations for both cases. It is important to keep in mind that each result involved a new training of the CNN model based on the chosen subset of modalities.

\subsubsection{Cockpit Use Case}

In the cockpit use case, the model trained with only finger modality performs the best amongst the models trained with single modalities. It has an MAD of 6.1° which is lower than the ones obtained using only eye or only head, even though the standard deviation of finger is relatively high (see Figure \ref{fig:MAD}). By adding eye pose or head pose, the MAD is reduced to 3.7° or 2.7°, respectively, thereby enhancing the precision. Fusion of all three modalities has the best outcome with an MAD of 2.5°. The same trend can be seen in hit rates. Finger has a hit rate of 75\%, which is the highest hit rate amongst the three modalities. The combination of all three modalities increases the hit rate to about 87\% (see Figure \ref{fig:hit_rate}).

Finger modality exhibits the highest precision (i.e., an MAD of 6.1°). This is due to the small distance of the AOIs from the finger tip as the AOIs are close by to the driver's hands. Eye gaze being less precise is counter intuitive at first glance. However, a deeper analysis of the recorded videos of the drivers from the additional four cameras discussed in Section \ref{sec:apparatus} revealed that the drivers were mostly looking downwards and in some cases, the eyelids covered the pupils partially which caused erroneous tracking of the gaze. This was especially true for AOIs 1, 2 and 3 which lie near the gearbox. Because of the erroneous tracking and the volatile nature of eye gaze, its MAD of 7.2° is the largest in the cockpit use case, with a Std.AD of 8.6°. It is interesting to see that despite the larger MAD of gaze modality compared to head, the hit rate for gaze modality of 56\% is slightly higher than that for head modality. This indicates that  with gaze, the predictions hit the target AOI slightly more, but on the other hand, predictions using only gaze have more outliers compared to head, which cause the Std.AD to be relatively high.

Even though finger has the most hits amongst all three modalities, it has a relatively high Std.AD, indicating many outliers (i.e., predictions with very large error). Upon careful analysis of the sensor data and looking at the recorded videos from the additional cameras, we were able to identify a few reasons for the outliers. The main reason in the cockpit use case was the driver's left hand pointing towards AOI 10 and 11. Since these two lie below the steering wheel, far from the field-of-view of the GCS, the tracking of the left hand often had erroneous measured data, or there was no tracking of the left hand. Among other reasons, is different pointing direction for the same AOI, e.g. for AOI 9 the majority of the users pointed with right hand but there are a few samples with the left hand as well. Since the AOI 9 is so close to the driver, the change of hand causes a huge difference in the angle of pointing direction. This is because the finger position and direction were different for each event w.r.t the ground truth, which is always kept the same for consistency. 

\subsubsection{Environment Use Case}\label{sec:results_environment}
In this case, finger modality has the least precision (i.e., the highest MAD at 19.9°), while eye has the best precision among the three modalities with an MAD of 9.3°. Finger modality has a very high standard deviation as well, i.e. an Std.AD of 21.7 (see Figure \ref{fig:MAD}). It was observed that when using only the finger modality, the predictions had many outliers with very large errors which cause the larger value of MAD as well as Std.AD. In addition to outliers, due to the relatively larger size of POIs in comparison with the size of AOIs,  the participants' pointing directions have a large variance as they were free to choose any place on the POI to point. This is one of the reasons why a relatively larger dataset was required for an adequate prediction precision in environment use case as compared to cockpit use case. However, the main reason for the finger modality being least precise was the use of left hand for pointing to the left side, which resulted in the hand dropping out of the field of view of the gesture camera causing partial unavailability of the finger pose. This happens in about 20\% of the referencing events. The majority of the outliers lie in the car poses 2 and 3, each having about 49\% outliers, while car pose 1 and 4 only had about 1\% outliers. In car pose 3, there are about 30\% pointing events with the left hand use, while other positions each have about 10\% with the left hand.

Looking at the hit rate in the environment use case in Figure \ref{fig:hit_rate}, eye modality has the highest hit rate at 61\% while finger modality has the lowest hit rate at 38\%. These results are almost contrary to the cockpit use case. Amongst single modalities, eye has the highest hit rate in environment use case and the lowest precision in the cockpit use case (see Figure \ref{fig:hit_rate}), while finger has the lowest hit rate as well as precision. Whereas in the cockpit use case, finger has the highest hit rate and precision, while eye has the lowest precision. Therefore, to tackle both use cases, the use of only uni-modal input would not be an optimal choice as the above-mentioned results reveal that two different modalities perform best in the two use cases.

Furthermore, the fusion of all modalities increases the hit rate up to 68\% and reduces the MAD to 7°. However, the Std.AD of 8.6° obtained when fusing all three modalities is still relatively high in the environment use case. This is because, despite the relatively high availability of eye compared to finger, the eye gaze features are often missing especially when looking to the far right or the far left, such as in car pose 2 or car pose 3, respectively. Occlusion of the face occurs often as well when the pointing arm appears in front of the face, thereby occluding the head. Nevertheless, the results obtained from fusing all three modalities show a clear improvement over uni-modal inputs in terms of hit rate as well as MAD.

\begin{figure*}[ht]
       \centering
  \includegraphics[trim=0 0 0 0,clip, width=\textwidth]{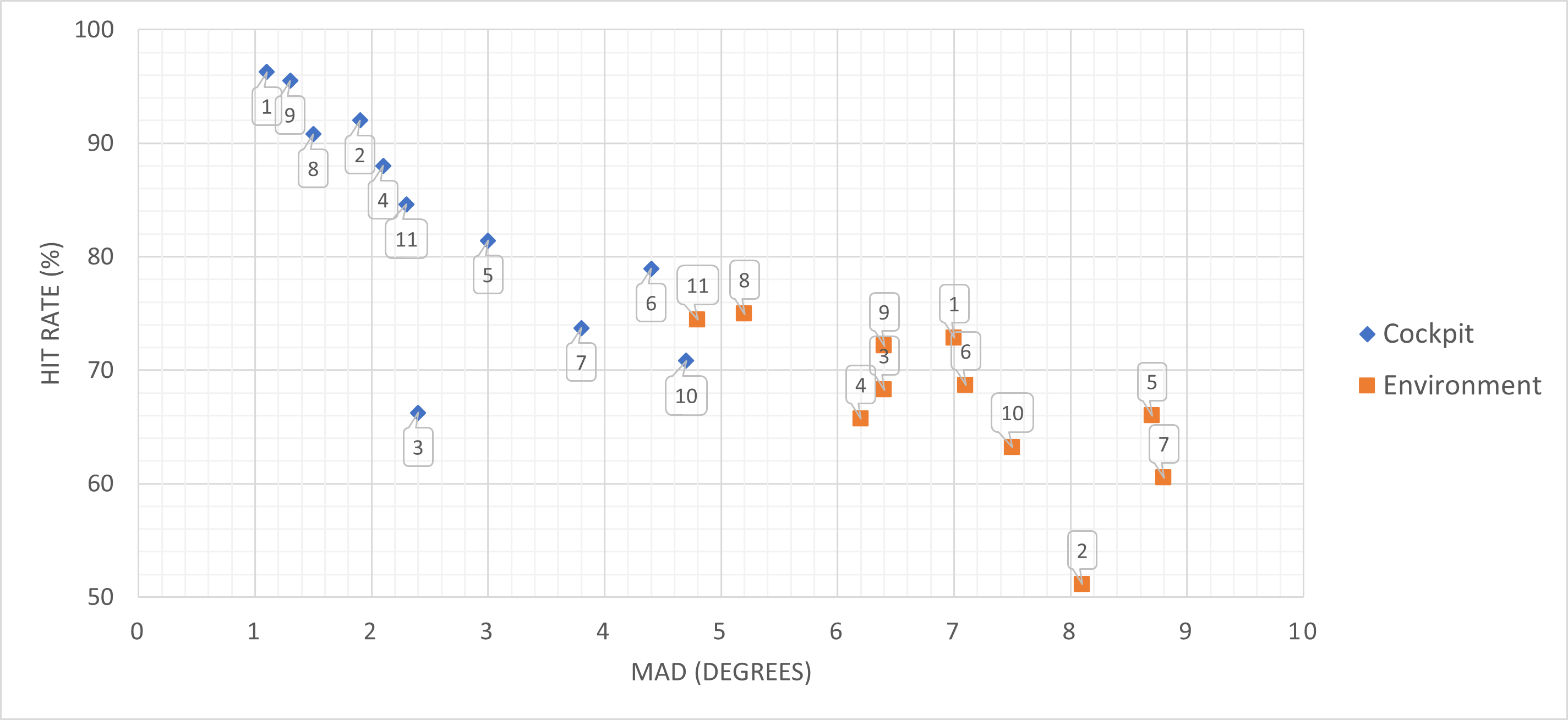}
      \centering
  \caption{Drivers' specific results in terms of MAD and hit rate using the fusion of all three modalities (head, eye and finger)}
  \label{fig:drivers}
\end{figure*}

\subsection{Driver Specific Results}

The results pertaining to specific driver behaviors are shown in Figure \ref{fig:drivers} for all 11 drivers. The precision and hit rate obtained for each driver in both cases have a large variance. An interesting outcome that we observe here is that some drivers have good pointing precision as well as hit rate in one use case, but not in the other. For example, driver '2' has a relatively high hit rate and low MAD, i.e., high precision, in the cockpit use case (shown with blue diamond markers), but has the lowest hit rate and a relatively low precision in the environment use case (shown with orange square markers).  Similarly, driver '3' has the lowest hit rate in cockpit use case, but has slightly above average hit rate and precision in the environment case. Drivers '1' and '9' appear to have the best hit rate and precision in cockpit referencing, while drivers '11' and '8' appear to have the best hit rate and precision for referencing in the environment use case. With this, we can conclude that drivers behave differently in different cases. More specifically, drivers have different strengths and weaknesses as to how accurate they are in employing the various modalities, and an extensive study of both cases is indeed necessary.

\subsection{Cross-dataset Learning} \label{sec:cross_dataset}

\begin{figure}[t]
\begin{subfigure}{.478\textwidth}
 \centering
  \includegraphics[trim=0 0 0 0,clip, width=\textwidth]{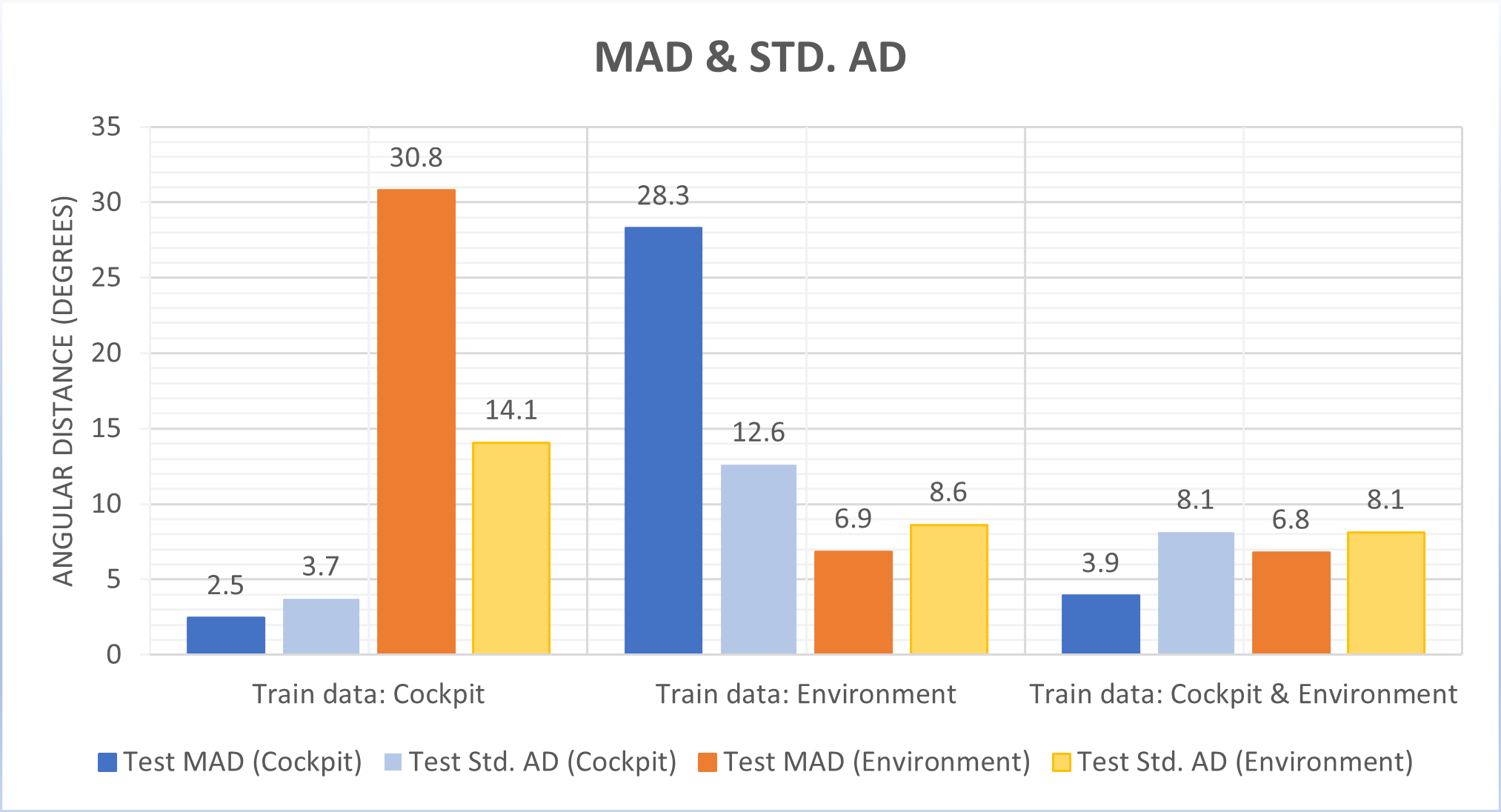}
  \caption{MAD and Std.AD of cross-dataset learning}
  \label{fig:cross_dataset_mad}
\end{subfigure} 
\begin{subfigure}{.478\textwidth}
 \centering
  \includegraphics[trim=0 0 0 0,clip, width=\textwidth]{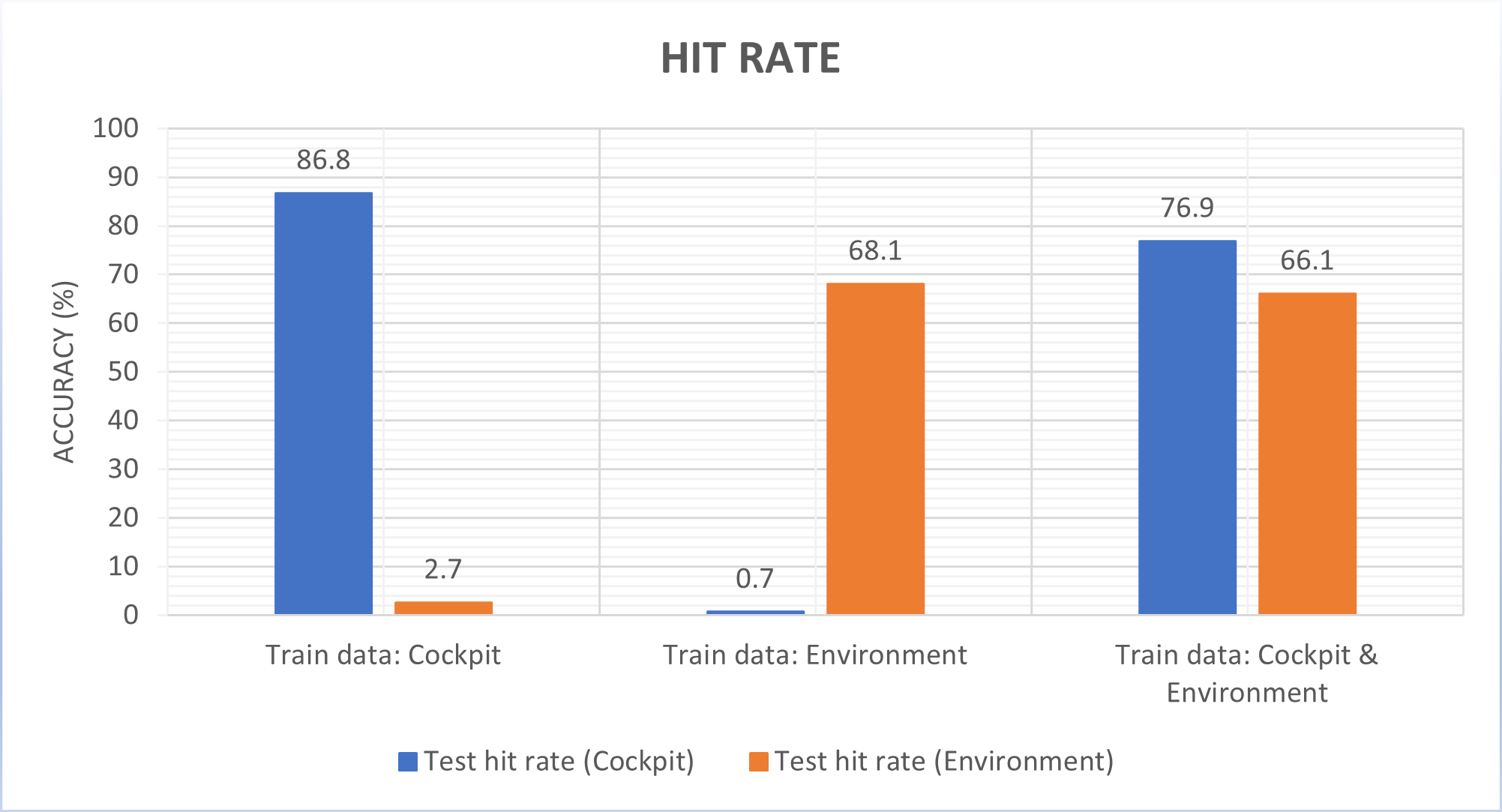}
  \caption{Hit rate of cross-dataset learning}
  \label{fig:cross_dataset_hit}
\end{subfigure}
  \caption{Cross-dataset learning MAD, Std.AD and hit rate using the fusion of all three modalities (head, eye and finger)}
  \label{fig:cross_dataset}
\end{figure}

In the previous sections, we either used only cockpit data for training as well as testing or we used only environment data for training as well as testing. In this section, in order to show the differences between the two use cases, we conducted multiple tests with cross-dataset learning of the CNN model. This means for testing on environment use case, the model training uses the cockpit dataset only and model testing uses the environment data only, and vice versa for testing on cockpit use case. The results from the cross-dataset learning are shown in Figure \ref{fig:cross_dataset}. 

The MAD and Std.AD of the model trained on cockpit dataset and tested on the environment dataset are 30.8° and 14.1°, respectively (see Figure \ref{fig:cross_dataset_mad}), which are significantly higher than when the model was trained and tested on environment data only. A hit rate of only 2.7\% was achieved in this test (see Figure \ref{fig:cross_dataset_hit}). Similarly, using only the environment dataset for model training, and testing on cockpit dataset, an MAD and Std.Ad of 28.3° and 12.6°, respectively, were achieved, while only 0.7\% of the events hit the target AOI. Despite the fact that the users simply point to either an AOI or POI in both uses cases, the results obtained from cross-dataset learning indicate that one model can not be generalized for both use cases in an optimal manner. This might be caused  by the unavailability of certain modalities in either use case. For example, we observed that finger modality has the highest availability in the cockpit use case with the highest hit rate as well, while head modality had the highest availability in the environment use case and eye modality had the highest hit rate. Consequently, a model trained using only inside vehicle data will rely more on the finger modality, which we have shown to be less precise than gaze in the outside vehicle use case, resulting in decreased precision for POIs. Therefore, in order to have an application with deictic referencing, it is vital to have the appropriate variation of pointing directions in the dataset that will be used for training.

Furthermore, we conducted tests using all data (i.e., using both cockpit data and environment data simultaneously for training). The test MAD and test Std.AD for cockpit use case were 3.9° an 8.1°, respectively, which is 1.4° larger in MAD and 4.4° larger in the Std.AD than when we used only cockpit data for training. The hit rate decreased from 86.8\% to 76.9\%. When testing on the environment data with the model trained using both datasets, no significant changes were seen. This might be induced by the higher ratio of environment data compared to cockpit data.

\section{Conclusion}
In this paper, we analyzed and studied features from three modalities, eye-gaze, head and finger, to determine the driver's referenced object, while using speech as a trigger. The experiments were divided into two types, pointing to objects inside the vehicle and pointing to objects outside the vehicle. For the objects inside the vehicle, finger pointing was observed to be the dominant modality, whereas, for the objects outside, gaze was the dominant modality  amongst the three. This shows that there is not a single  modality that would be optimal for both types of pointing. Rather, as the sensors do not offer 100\% tracking availability for single modalities because of multiple factors such as occlusion or movement out of view, there is a need for multimodal fusion for improved recognition of the referenced direction as well as for better generalization for different use cases.

Therefore, we propose a 2-stage CNN based multimodal fusion architecture to initially determine whether the driver's referenced object lies inside or outside the vehicle. In the second step, based on the recognized use case type, the appropriately trained model for fusion of the modalities is applied to better estimate the pointing direction. We successfully identified the placement of the referenced object to be inside or outside the vehicle with an accuracy of 98.6\%. The fusion of all three modalities has been shown to outperform individual modalities, in terms of both the mean angular distance as well as the hit rate. Referencing interior objects reveals to be more precise than for the exterior objects. The hit rate for the interior objects is also shown to be greater than the hit rate of exterior objects mainly due to the shorter distance of the interior objects to the driver's hands.
Furthermore, we compared mean angular distance and hit rate of the drivers in both cases, and concluded that drivers' referencing behavior is different in the two cases. 
In addition to this, we compared cross-dataset performances in the two use cases and illustrated that one use case can not produce sufficiently good results in the other because the two use cases in fact exhibit different limitations and conditions. Simultaneous use of all data for a generalized approach may be one solution, however, this results in a slight reduction in precision (i.e., increase in mean angular distance) as well as hit rate for inside-vehicle objects.

In general, our paper provides a novel application of natural user interaction for driver assistance systems exploiting the inter-dependencies between multiple modalities. This paves new ways for further work in recognizing the driver's referencing intent which would include both the referenced object as well as the action to be taken, allowing a more natural user experience.

\begin{acks}
We are grateful to Ovidiu Bulzan and Stefan Schubert (BMW Group, Munich) for their contributions to the experiment design and apparatus setup. We would also like to thank Steven Rohrhirsch, Tobias Brosch and Benoit Diotte (BMW Car IT GmbH, Ulm) for their support in data extraction and pre-processing steps.
\end{acks}

\bibliographystyle{ACM-Reference-Format}
\bibliography{main}


\begin{thebibliography}{38}

%%% the Web address.

\ifx \showCODEN    \undefined \def \showCODEN     #1{\unskip}     \fi
\ifx \showDOI      \undefined \def \showDOI       #1{#1}\fi
\ifx \showISBNx    \undefined \def \showISBNx     #1{\unskip}     \fi
\ifx \showISBNxiii \undefined \def \showISBNxiii  #1{\unskip}     \fi
\ifx \showISSN     \undefined \def \showISSN      #1{\unskip}     \fi
\ifx \showLCCN     \undefined \def \showLCCN      #1{\unskip}     \fi
\ifx \shownote     \undefined \def \shownote      #1{#1}          \fi
\ifx \showarticletitle \undefined \def \showarticletitle #1{#1}   \fi
\ifx \showURL      \undefined \def \showURL       {\relax}        \fi
% invisible to TeX
\providecommand\bibfield[2]{#2}
\providecommand\bibinfo[2]{#2}
\providecommand\natexlab[1]{#1}
\providecommand\showeprint[2][]{arXiv:#2}

\bibitem[\protect\citeauthoryear{Aftab, von~der Beeck, and Feld}{Aftab
  et~al\mbox{.}}{2020}]%
        {aftab2020you}
\bibfield{author}{\bibinfo{person}{Abdul~Rafey Aftab}, \bibinfo{person}{Michael
  von~der Beeck}, {and} \bibinfo{person}{Michael Feld}.}
  \bibinfo{year}{2020}\natexlab{}.
\newblock \showarticletitle{You Have a Point There: Object Selection Inside an
  Automobile Using Gaze, Head Pose and Finger Pointing}. In
  \bibinfo{booktitle}{\emph{Proceedings of the 2020 International Conference on
  Multimodal Interaction}}. \bibinfo{pages}{595--603}.
\newblock


\bibitem[\protect\citeauthoryear{Aftab, von~der Beeck, Rohrhirsch, Diotte, and
  Feld}{Aftab et~al\mbox{.}}{2021}]%
        {aftab2021multimodal}
\bibfield{author}{\bibinfo{person}{Abdul~Rafey Aftab}, \bibinfo{person}{Michael
  von~der Beeck}, \bibinfo{person}{Steven Rohrhirsch}, \bibinfo{person}{Benoit
  Diotte}, {and} \bibinfo{person}{Michael Feld}.}
  \bibinfo{year}{2021}\natexlab{}.
\newblock \showarticletitle{Multimodal Fusion Using Deep Learning Applied to
  Driver's Referencing of Outside-Vehicle Objects}. In
  \bibinfo{booktitle}{\emph{2021 IEEE Intelligent Vehicles Symposium (IV)}}.
  IEEE, \bibinfo{pages}{1108--1115}.
\newblock


\bibitem[\protect\citeauthoryear{Ahmad, Murphy, Godsill, Langdon, and
  Hardy}{Ahmad et~al\mbox{.}}{2017}]%
        {ahmad2017intelligent}
\bibfield{author}{\bibinfo{person}{Bashar~I Ahmad},
  \bibinfo{person}{James~Kevin Murphy}, \bibinfo{person}{Simon Godsill},
  \bibinfo{person}{Patrick~M Langdon}, {and} \bibinfo{person}{Robery Hardy}.}
  \bibinfo{year}{2017}\natexlab{}.
\newblock \showarticletitle{Intelligent interactive displays in vehicles with
  intent prediction: A Bayesian framework}.
\newblock \bibinfo{journal}{\emph{IEEE Signal Processing Magazine}}
  \bibinfo{volume}{34}, \bibinfo{number}{2} (\bibinfo{year}{2017}),
  \bibinfo{pages}{82--94}.
\newblock


\bibitem[\protect\citeauthoryear{Akkil and Isokoski}{Akkil and
  Isokoski}{2016}]%
        {akkil2016accuracy}
\bibfield{author}{\bibinfo{person}{Deepak Akkil} {and} \bibinfo{person}{Poika
  Isokoski}.} \bibinfo{year}{2016}\natexlab{}.
\newblock \showarticletitle{Accuracy of interpreting pointing gestures in
  egocentric view}. In \bibinfo{booktitle}{\emph{Proceedings of the 2016 ACM
  International Joint Conference on Pervasive and Ubiquitous Computing}}.
  \bibinfo{pages}{262--273}.
\newblock


\bibitem[\protect\citeauthoryear{Bolt}{Bolt}{1980}]%
        {bolt1980put}
\bibfield{author}{\bibinfo{person}{Richard~A Bolt}.}
  \bibinfo{year}{1980}\natexlab{}.
\newblock \showarticletitle{“Put-that-there” Voice and gesture at the
  graphics interface}. In \bibinfo{booktitle}{\emph{Proceedings of the 7th
  annual conference on Computer graphics and interactive techniques}}.
  \bibinfo{pages}{262--270}.
\newblock


\bibitem[\protect\citeauthoryear{Brand, B{\"u}chele, and
  Meschtscherjakov}{Brand et~al\mbox{.}}{2016}]%
        {brand2016pointing}
\bibfield{author}{\bibinfo{person}{Daniel Brand}, \bibinfo{person}{Kevin
  B{\"u}chele}, {and} \bibinfo{person}{Alexander Meschtscherjakov}.}
  \bibinfo{year}{2016}\natexlab{}.
\newblock \showarticletitle{{Pointing at the HUD: Gesture interaction using a
  leap motion}}. In \bibinfo{booktitle}{\emph{Adjunct Proceedings of the 8th
  International Conference on Automotive User Interfaces and Interactive
  Vehicular Applications}}. \bibinfo{pages}{167--172}.
\newblock


\bibitem[\protect\citeauthoryear{Chatterjee, Xiao, and Harrison}{Chatterjee
  et~al\mbox{.}}{2015}]%
        {chatterjee2015gaze+}
\bibfield{author}{\bibinfo{person}{Ishan Chatterjee}, \bibinfo{person}{Robert
  Xiao}, {and} \bibinfo{person}{Chris Harrison}.}
  \bibinfo{year}{2015}\natexlab{}.
\newblock \showarticletitle{Gaze+ gesture: Expressive, precise and targeted
  free-space interactions}. In \bibinfo{booktitle}{\emph{Proceedings of the
  2015 ACM on International Conference on Multimodal Interaction}}.
  \bibinfo{pages}{131--138}.
\newblock


\bibitem[\protect\citeauthoryear{Chen and Jin}{Chen and Jin}{2016}]%
        {chen2016multi}
\bibfield{author}{\bibinfo{person}{Shizhe Chen} {and} \bibinfo{person}{Qin
  Jin}.} \bibinfo{year}{2016}\natexlab{}.
\newblock \showarticletitle{Multi-modal conditional attention fusion for
  dimensional emotion prediction}. In \bibinfo{booktitle}{\emph{Proceedings of
  the 24th ACM international conference on Multimedia}}.
  \bibinfo{pages}{571--575}.
\newblock


\bibitem[\protect\citeauthoryear{Esteban, Starr, Willetts, Hannah, and
  Bryanston-Cross}{Esteban et~al\mbox{.}}{2005}]%
        {esteban2005review}
\bibfield{author}{\bibinfo{person}{Jaime Esteban}, \bibinfo{person}{Andrew
  Starr}, \bibinfo{person}{Robert Willetts}, \bibinfo{person}{Paul Hannah},
  {and} \bibinfo{person}{Peter Bryanston-Cross}.}
  \bibinfo{year}{2005}\natexlab{}.
\newblock \showarticletitle{A review of data fusion models and architectures:
  towards engineering guidelines}.
\newblock \bibinfo{journal}{\emph{Neural Computing \& Applications}}
  \bibinfo{volume}{14}, \bibinfo{number}{4} (\bibinfo{year}{2005}),
  \bibinfo{pages}{273--281}.
\newblock


\bibitem[\protect\citeauthoryear{Fujimura, Xu, Tran, Bhandari, and
  Ng-Thow-Hing}{Fujimura et~al\mbox{.}}{2013}]%
        {fujimura2013driver}
\bibfield{author}{\bibinfo{person}{Kikuo Fujimura}, \bibinfo{person}{Lijie Xu},
  \bibinfo{person}{Cuong Tran}, \bibinfo{person}{Rishabh Bhandari}, {and}
  \bibinfo{person}{Victor Ng-Thow-Hing}.} \bibinfo{year}{2013}\natexlab{}.
\newblock \showarticletitle{Driver queries using wheel-constrained finger
  pointing and 3-D head-up display visual feedback}. In
  \bibinfo{booktitle}{\emph{Proceedings of the 5th International Conference on
  Automotive User Interfaces and Interactive Vehicular Applications}}.
  \bibinfo{pages}{56--62}.
\newblock


\bibitem[\protect\citeauthoryear{Gomaa, Reyes, Alles, Rupp, and Feld}{Gomaa
  et~al\mbox{.}}{2020}]%
        {gomaa2020studying}
\bibfield{author}{\bibinfo{person}{Amr Gomaa}, \bibinfo{person}{Guillermo
  Reyes}, \bibinfo{person}{Alexandra Alles}, \bibinfo{person}{Lydia Rupp},
  {and} \bibinfo{person}{Michael Feld}.} \bibinfo{year}{2020}\natexlab{}.
\newblock \showarticletitle{Studying Person-Specific Pointing and Gaze Behavior
  for Multimodal Referencing of Outside Objects from a Moving Vehicle}. In
  \bibinfo{booktitle}{\emph{Proceedings of the 2020 International Conference on
  Multimodal Interaction}}. \bibinfo{pages}{501--509}.
\newblock


\bibitem[\protect\citeauthoryear{Gomaa, Reyes, and Feld}{Gomaa
  et~al\mbox{.}}{2021}]%
        {gomaa2021ml}
\bibfield{author}{\bibinfo{person}{Amr Gomaa}, \bibinfo{person}{Guillermo
  Reyes}, {and} \bibinfo{person}{Michael Feld}.}
  \bibinfo{year}{2021}\natexlab{}.
\newblock \showarticletitle{ML-PersRef: A Machine Learning-based Personalized
  Multimodal Fusion Approach for Referencing Outside Objects From a Moving
  Vehicle}. In \bibinfo{booktitle}{\emph{Proceedings of the 2021 International
  Conference on Multimodal Interaction}} (Montreal, Canada)
  \emph{(\bibinfo{series}{ICMI '21})}. \bibinfo{publisher}{Association for
  Computing Machinery}, \bibinfo{address}{New York, NY, USA},
  \bibinfo{numpages}{9}~pages.
\newblock
\showISBNx{9781450384810}
\urldef\tempurl%
\url{https://doi.org/10.1145/3462244.3479910}
\showDOI{\tempurl}


\bibitem[\protect\citeauthoryear{Hild, Peinsipp-Byma, Voit, and Beyerer}{Hild
  et~al\mbox{.}}{2019}]%
        {hild2019suggesting}
\bibfield{author}{\bibinfo{person}{Jutta Hild}, \bibinfo{person}{Elisabeth
  Peinsipp-Byma}, \bibinfo{person}{Michael Voit}, {and}
  \bibinfo{person}{J{\"u}rgen Beyerer}.} \bibinfo{year}{2019}\natexlab{}.
\newblock \showarticletitle{Suggesting Gaze-based Selection for Surveillance
  Applications}. In \bibinfo{booktitle}{\emph{2019 16th IEEE International
  Conference on Advanced Video and Signal Based Surveillance (AVSS)}}. IEEE,
  \bibinfo{pages}{1--8}.
\newblock


\bibitem[\protect\citeauthoryear{{ISO 8855: 2011 (E)}}{{ISO 8855: 2011
  (E)}}{2011}]%
        {iso2011road}
\bibfield{author}{\bibinfo{person}{{ISO 8855: 2011 (E)}}.}
  \bibinfo{year}{2011}\natexlab{}.
\newblock \showarticletitle{{Road vehicles—Vehicle dynamics and
  road—holding ability—Vocabulary}}.
\newblock \bibinfo{journal}{\emph{International Organization for
  Standardization Geneva}} (\bibinfo{year}{2011}).
\newblock


\bibitem[\protect\citeauthoryear{Ji and Yang}{Ji and Yang}{2002}]%
        {ji2002real}
\bibfield{author}{\bibinfo{person}{Qiang Ji} {and} \bibinfo{person}{Xiaojie
  Yang}.} \bibinfo{year}{2002}\natexlab{}.
\newblock \showarticletitle{Real-time eye, gaze, and face pose tracking for
  monitoring driver vigilance}.
\newblock \bibinfo{journal}{\emph{Real-time imaging}} \bibinfo{volume}{8},
  \bibinfo{number}{5} (\bibinfo{year}{2002}), \bibinfo{pages}{357--377}.
\newblock


\bibitem[\protect\citeauthoryear{Kang, Kim, Han, and Kim}{Kang
  et~al\mbox{.}}{2015}]%
        {kang2015you}
\bibfield{author}{\bibinfo{person}{Shinjae Kang}, \bibinfo{person}{Byungjo
  Kim}, \bibinfo{person}{Sangrok Han}, {and} \bibinfo{person}{Hyogon Kim}.}
  \bibinfo{year}{2015}\natexlab{}.
\newblock \showarticletitle{Do you see what I see: towards a gaze-based
  surroundings query processing system}. In
  \bibinfo{booktitle}{\emph{Proceedings of the 7th International Conference on
  Automotive User Interfaces and Interactive Vehicular Applications}}.
  \bibinfo{pages}{93--100}.
\newblock


\bibitem[\protect\citeauthoryear{Kim and Misu}{Kim and Misu}{2014}]%
        {kim2014identification}
\bibfield{author}{\bibinfo{person}{Young-Ho Kim} {and}
  \bibinfo{person}{Teruhisa Misu}.} \bibinfo{year}{2014}\natexlab{}.
\newblock \showarticletitle{Identification of the driver's interest point using
  a head pose trajectory for situated dialog systems}. In
  \bibinfo{booktitle}{\emph{Proceedings of the 16th International Conference on
  Multimodal Interaction}}. \bibinfo{pages}{92--95}.
\newblock


\bibitem[\protect\citeauthoryear{Liu, Shen, Lakshminarasimhan, Liang, Zadeh,
  and Morency}{Liu et~al\mbox{.}}{2018}]%
        {liu2018efficient}
\bibfield{author}{\bibinfo{person}{Zhun Liu}, \bibinfo{person}{Ying Shen},
  \bibinfo{person}{Varun~Bharadhwaj Lakshminarasimhan},
  \bibinfo{person}{Paul~Pu Liang}, \bibinfo{person}{Amir Zadeh}, {and}
  \bibinfo{person}{Louis-Philippe Morency}.} \bibinfo{year}{2018}\natexlab{}.
\newblock \showarticletitle{Efficient low-rank multimodal fusion with
  modality-specific factors}.
\newblock \bibinfo{journal}{\emph{arXiv preprint arXiv:1806.00064}}
  (\bibinfo{year}{2018}).
\newblock


\bibitem[\protect\citeauthoryear{Maglio, Matlock, Campbell, Zhai, and
  Smith}{Maglio et~al\mbox{.}}{2000}]%
        {maglio2000gaze}
\bibfield{author}{\bibinfo{person}{Paul~P Maglio}, \bibinfo{person}{Teenie
  Matlock}, \bibinfo{person}{Christopher~S Campbell}, \bibinfo{person}{Shumin
  Zhai}, {and} \bibinfo{person}{Barton~A Smith}.}
  \bibinfo{year}{2000}\natexlab{}.
\newblock \showarticletitle{Gaze and speech in attentive user interfaces}. In
  \bibinfo{booktitle}{\emph{International Conference on Multimodal
  Interfaces}}. Springer, \bibinfo{pages}{1--7}.
\newblock


\bibitem[\protect\citeauthoryear{Mayer, Schwind, Schweigert, and Henze}{Mayer
  et~al\mbox{.}}{2018}]%
        {mayer2018effect}
\bibfield{author}{\bibinfo{person}{Sven Mayer}, \bibinfo{person}{Valentin
  Schwind}, \bibinfo{person}{Robin Schweigert}, {and} \bibinfo{person}{Niels
  Henze}.} \bibinfo{year}{2018}\natexlab{}.
\newblock \showarticletitle{The effect of offset correction and cursor on
  mid-air pointing in real and virtual environments}. In
  \bibinfo{booktitle}{\emph{Proceedings of the 2018 CHI Conference on Human
  Factors in Computing Systems}}. \bibinfo{pages}{1--13}.
\newblock


\bibitem[\protect\citeauthoryear{Meng, Jing, Yan, and Pedrycz}{Meng
  et~al\mbox{.}}{2020}]%
        {meng2020survey}
\bibfield{author}{\bibinfo{person}{Tong Meng}, \bibinfo{person}{Xuyang Jing},
  \bibinfo{person}{Zheng Yan}, {and} \bibinfo{person}{Witold Pedrycz}.}
  \bibinfo{year}{2020}\natexlab{}.
\newblock \showarticletitle{A survey on machine learning for data fusion}.
\newblock \bibinfo{journal}{\emph{Information Fusion}}  \bibinfo{volume}{57}
  (\bibinfo{year}{2020}), \bibinfo{pages}{115--129}.
\newblock


\bibitem[\protect\citeauthoryear{Misu, Raux, Gupta, and Lane}{Misu
  et~al\mbox{.}}{2014}]%
        {misu2014situated}
\bibfield{author}{\bibinfo{person}{Teruhisa Misu}, \bibinfo{person}{Antoine
  Raux}, \bibinfo{person}{Rakesh Gupta}, {and} \bibinfo{person}{Ian Lane}.}
  \bibinfo{year}{2014}\natexlab{}.
\newblock \showarticletitle{Situated language understanding at 25 miles per
  hour}. In \bibinfo{booktitle}{\emph{Proceedings of the 15th Annual Meeting of
  the Special Interest Group on Discourse and Dialogue (SIGDIAL)}}.
  \bibinfo{pages}{22--31}.
\newblock


\bibitem[\protect\citeauthoryear{Mitrevska, Moniri, Ne{\ss}elrath, Schwartz,
  Feld, K{\"o}rber, Deru, and M{\"u}ller}{Mitrevska et~al\mbox{.}}{2015}]%
        {mitrevska2015siam}
\bibfield{author}{\bibinfo{person}{Monika Mitrevska},
  \bibinfo{person}{Mohammad~Mehdi Moniri}, \bibinfo{person}{Robert
  Ne{\ss}elrath}, \bibinfo{person}{Tim Schwartz}, \bibinfo{person}{Michael
  Feld}, \bibinfo{person}{Yannick K{\"o}rber}, \bibinfo{person}{Matthieu Deru},
  {and} \bibinfo{person}{Christian M{\"u}ller}.}
  \bibinfo{year}{2015}\natexlab{}.
\newblock \showarticletitle{SiAM-situation-adaptive multimodal interaction for
  innovative mobility concepts of the future}. In
  \bibinfo{booktitle}{\emph{2015 International Conference on Intelligent
  Environments}}. IEEE, \bibinfo{pages}{180--183}.
\newblock


\bibitem[\protect\citeauthoryear{Moniri and M{\"u}ller}{Moniri and
  M{\"u}ller}{2012}]%
        {moniri2012multimodal}
\bibfield{author}{\bibinfo{person}{Mohammad~Mehdi Moniri} {and}
  \bibinfo{person}{Christian M{\"u}ller}.} \bibinfo{year}{2012}\natexlab{}.
\newblock \showarticletitle{Multimodal reference resolution for mobile spatial
  interaction in urban environments}. In \bibinfo{booktitle}{\emph{Proceedings
  of the 4th International Conference on Automotive User Interfaces and
  Interactive Vehicular Applications}}. \bibinfo{pages}{241--248}.
\newblock


\bibitem[\protect\citeauthoryear{Mukherjee and Robertson}{Mukherjee and
  Robertson}{2015}]%
        {mukherjee2015deep}
\bibfield{author}{\bibinfo{person}{Sankha~S Mukherjee} {and}
  \bibinfo{person}{Neil~Martin Robertson}.} \bibinfo{year}{2015}\natexlab{}.
\newblock \showarticletitle{Deep head pose: Gaze-direction estimation in
  multimodal video}.
\newblock \bibinfo{journal}{\emph{IEEE Transactions on Multimedia}}
  \bibinfo{volume}{17}, \bibinfo{number}{11} (\bibinfo{year}{2015}),
  \bibinfo{pages}{2094--2107}.
\newblock


\bibitem[\protect\citeauthoryear{M{\"u}ller and Weinberg}{M{\"u}ller and
  Weinberg}{2011}]%
        {muller2011multimodal}
\bibfield{author}{\bibinfo{person}{Christian M{\"u}ller} {and}
  \bibinfo{person}{Garrett Weinberg}.} \bibinfo{year}{2011}\natexlab{}.
\newblock \showarticletitle{Multimodal input in the car, today and tomorrow}.
\newblock \bibinfo{journal}{\emph{IEEE MultiMedia}} \bibinfo{volume}{18},
  \bibinfo{number}{1} (\bibinfo{year}{2011}), \bibinfo{pages}{98--103}.
\newblock


\bibitem[\protect\citeauthoryear{Ne{\ss}elrath, Moniri, and Feld}{Ne{\ss}elrath
  et~al\mbox{.}}{2016}]%
        {nesselrath2016combining}
\bibfield{author}{\bibinfo{person}{Robert Ne{\ss}elrath},
  \bibinfo{person}{Mohammad~Mehdi Moniri}, {and} \bibinfo{person}{Michael
  Feld}.} \bibinfo{year}{2016}\natexlab{}.
\newblock \showarticletitle{Combining speech, gaze, and micro-gestures for the
  multimodal control of in-car functions}. In \bibinfo{booktitle}{\emph{2016
  12th International Conference on Intelligent Environments (IE)}}. IEEE,
  \bibinfo{pages}{190--193}.
\newblock


\bibitem[\protect\citeauthoryear{Ngiam, Khosla, Kim, Nam, Lee, and Ng}{Ngiam
  et~al\mbox{.}}{2011}]%
        {ngiam2011multimodal}
\bibfield{author}{\bibinfo{person}{Jiquan Ngiam}, \bibinfo{person}{Aditya
  Khosla}, \bibinfo{person}{Mingyu Kim}, \bibinfo{person}{Juhan Nam},
  \bibinfo{person}{Honglak Lee}, {and} \bibinfo{person}{Andrew~Y Ng}.}
  \bibinfo{year}{2011}\natexlab{}.
\newblock \showarticletitle{Multimodal deep learning}.
\newblock  (\bibinfo{year}{2011}).
\newblock


\bibitem[\protect\citeauthoryear{Ohn-Bar, Martin, Tawari, and Trivedi}{Ohn-Bar
  et~al\mbox{.}}{2014}]%
        {ohn2014head}
\bibfield{author}{\bibinfo{person}{Eshed Ohn-Bar}, \bibinfo{person}{Sujitha
  Martin}, \bibinfo{person}{Ashish Tawari}, {and} \bibinfo{person}{Mohan~M
  Trivedi}.} \bibinfo{year}{2014}\natexlab{}.
\newblock \showarticletitle{Head, eye, and hand patterns for driver activity
  recognition}. In \bibinfo{booktitle}{\emph{2014 22nd international conference
  on pattern recognition}}. IEEE, \bibinfo{pages}{660--665}.
\newblock


\bibitem[\protect\citeauthoryear{Pfleging, Schneegass, and Schmidt}{Pfleging
  et~al\mbox{.}}{2012}]%
        {pfleging2012multimodal}
\bibfield{author}{\bibinfo{person}{Bastian Pfleging}, \bibinfo{person}{Stefan
  Schneegass}, {and} \bibinfo{person}{Albrecht Schmidt}.}
  \bibinfo{year}{2012}\natexlab{}.
\newblock \showarticletitle{Multimodal interaction in the car: combining speech
  and gestures on the steering wheel}. In \bibinfo{booktitle}{\emph{Proceedings
  of the 4th International Conference on Automotive User Interfaces and
  Interactive Vehicular Applications}}. \bibinfo{pages}{155--162}.
\newblock


\bibitem[\protect\citeauthoryear{Roider and Gross}{Roider and Gross}{2018}]%
        {roider2018see}
\bibfield{author}{\bibinfo{person}{Florian Roider} {and} \bibinfo{person}{Tom
  Gross}.} \bibinfo{year}{2018}\natexlab{}.
\newblock \showarticletitle{I See Your Point: Integrating Gaze to Enhance
  Pointing Gesture Accuracy While Driving}. In
  \bibinfo{booktitle}{\emph{Proceedings of the 10th International Conference on
  Automotive User Interfaces and Interactive Vehicular Applications}}.
  \bibinfo{pages}{351--358}.
\newblock


\bibitem[\protect\citeauthoryear{Roider and Raab}{Roider and Raab}{2018}]%
        {roider2018implementation}
\bibfield{author}{\bibinfo{person}{Florian Roider} {and}
  \bibinfo{person}{Konstantin Raab}.} \bibinfo{year}{2018}\natexlab{}.
\newblock \showarticletitle{Implementation and Evaluation of Peripheral Light
  Feedback for Mid-Air Gesture Interaction in the Car}. In
  \bibinfo{booktitle}{\emph{2018 14th International Conference on Intelligent
  Environments (IE)}}. IEEE, \bibinfo{pages}{87--90}.
\newblock


\bibitem[\protect\citeauthoryear{R{\"u}melin, Marouane, and Butz}{R{\"u}melin
  et~al\mbox{.}}{2013}]%
        {rumelin2013free}
\bibfield{author}{\bibinfo{person}{Sonja R{\"u}melin}, \bibinfo{person}{Chadly
  Marouane}, {and} \bibinfo{person}{Andreas Butz}.}
  \bibinfo{year}{2013}\natexlab{}.
\newblock \showarticletitle{Free-hand pointing for identification and
  interaction with distant objects}. In \bibinfo{booktitle}{\emph{Proceedings
  of the 5th International Conference on Automotive User Interfaces and
  Interactive Vehicular Applications}}. \bibinfo{pages}{40--47}.
\newblock


\bibitem[\protect\citeauthoryear{Sauras-Perez, Gil, Gill, Pisu, and
  Taiber}{Sauras-Perez et~al\mbox{.}}{2017}]%
        {sauras2017voge}
\bibfield{author}{\bibinfo{person}{Pablo Sauras-Perez}, \bibinfo{person}{Andrea
  Gil}, \bibinfo{person}{Jasprit~Singh Gill}, \bibinfo{person}{Pierluigi Pisu},
  {and} \bibinfo{person}{Joachim Taiber}.} \bibinfo{year}{2017}\natexlab{}.
\newblock \bibinfo{booktitle}{\emph{VoGe: A Voice and Gesture System for
  Interacting with Autonomous Cars}}.
\newblock \bibinfo{type}{{T}echnical {R}eport}. \bibinfo{institution}{SAE
  Technical Paper}.
\newblock


\bibitem[\protect\citeauthoryear{Schweigert, Schwind, and Mayer}{Schweigert
  et~al\mbox{.}}{2019}]%
        {schweigert2019eyepointing}
\bibfield{author}{\bibinfo{person}{Robin Schweigert}, \bibinfo{person}{Valentin
  Schwind}, {and} \bibinfo{person}{Sven Mayer}.}
  \bibinfo{year}{2019}\natexlab{}.
\newblock \showarticletitle{EyePointing: A Gaze-Based Selection Technique}.
\newblock In \bibinfo{booktitle}{\emph{Proceedings of Mensch und Computer
  2019}}. \bibinfo{pages}{719--723}.
\newblock


\bibitem[\protect\citeauthoryear{Tscharn, Latoschik, L{\"o}ffler, and
  Hurtienne}{Tscharn et~al\mbox{.}}{2017}]%
        {tscharn2017stop}
\bibfield{author}{\bibinfo{person}{Robert Tscharn}, \bibinfo{person}{Marc~Erich
  Latoschik}, \bibinfo{person}{Diana L{\"o}ffler}, {and}
  \bibinfo{person}{J{\"o}rn Hurtienne}.} \bibinfo{year}{2017}\natexlab{}.
\newblock \showarticletitle{“Stop over there”: natural gesture and speech
  interaction for non-critical spontaneous intervention in autonomous driving}.
  In \bibinfo{booktitle}{\emph{Proceedings of the 19th acm international
  conference on multimodal interaction}}. \bibinfo{pages}{91--100}.
\newblock


\bibitem[\protect\citeauthoryear{Turk}{Turk}{2014}]%
        {turk2014multimodal}
\bibfield{author}{\bibinfo{person}{Matthew Turk}.}
  \bibinfo{year}{2014}\natexlab{}.
\newblock \showarticletitle{Multimodal interaction: A review}.
\newblock \bibinfo{journal}{\emph{Pattern Recognition Letters}}
  \bibinfo{volume}{36} (\bibinfo{year}{2014}), \bibinfo{pages}{189--195}.
\newblock


\bibitem[\protect\citeauthoryear{Wu, Lin, and Wei}{Wu et~al\mbox{.}}{2014}]%
        {wu2014survey}
\bibfield{author}{\bibinfo{person}{Chung-Hsien Wu}, \bibinfo{person}{Jen-Chun
  Lin}, {and} \bibinfo{person}{Wen-Li Wei}.} \bibinfo{year}{2014}\natexlab{}.
\newblock \showarticletitle{Survey on audiovisual emotion recognition:
  databases, features, and data fusion strategies}.
\newblock \bibinfo{journal}{\emph{APSIPA transactions on signal and information
  processing}}  \bibinfo{volume}{3} (\bibinfo{year}{2014}).
\newblock


\end{thebibliography}

\end{document}